\newif\ifrsb
\rsbfalse

\ifrsb
    \documentclass[openacc]{rstb} 
    \titlehead{Research}

\else
    \documentclass[english,reprint,aps,pra]{revtex4-1}
    \usepackage[unicode=true,bookmarks=false,breaklinks=false,pdfborder={0 0 1},colorlinks=false]{hyperref}
    \hypersetup{colorlinks,linkcolor=DarkBlue,citecolor=DarkBlue}
    \usepackage[svgnames,x11names,table]{xcolor}

    \usepackage{newtxtext}

    \usepackage[T1]{fontenc}
    \usepackage[utf8]{inputenc}
    \usepackage{amsmath}
    \usepackage{amssymb}

    \allowdisplaybreaks

    \makeatother

    \usepackage{babel}
\fi

\usepackage{graphicx}

\usepackage{xargs}[2008/03/08]

\usepackage{xcolor}

\usepackage{chemfig}

\usepackage{tabularray}
\global\long\def\makess#1{#1^{*}}%
\global\long\def\xx{x}%
\global\long\def\dil{\phi}%
\global\long\def\aa{\vec{a}}%
\global\long\def\MM{A}%
\global\long\def\MMi{\MM_{i}}%
\global\long\def\Amatrix{M}%
\global\long\def\xxss{\makess{\xx}}%
\global\long\def\epss{\makess{\sigma}}%
\global\long\def\yy{\vec{y}}%
\global\long\def\dilss{\dil}%

\global\long\def\fitness{f}%

\global\long\def\rrArg#1{\nu_{#1}}%
\global\long\def\rrMArg#1{\nu_{#1}^{-}}%
\global\long\def\vv{\rrArg k}%
\global\long\def\vvM{\rrMArg k}%

\global\long\def\ww{r}%
\global\long\def\wwM{r^{-}}%
\newcommandx\wwd[1][usedefault, addprefix=\global, 1=\dil]{r(#1)}%
\newcommandx\wwdM[1][usedefault, addprefix=\global, 1=\dil]{r^{-}(#1)}%
\global\long\def\dilcrit{\hat{\phi}}%
\global\long\def\ep{\sigma}%
\global\long\def\epK{\sigma_{k}}%

\global\long\def\msIx{k}%
\global\long\def\msIxx{l}%
\global\long\def\msNum{m}%
\newcommandx\JJ[1][usedefault, addprefix=\global, 1=]{J_{#1}}%
\global\long\def\rr{\rho}%

\global\long\def\kkM#1{r_{#1}^{-}(\dilss)}%
\global\long\def\kk#1{r_{#1}(\dilss)}%

\global\long\def\ccIx{j}%
\global\long\def\ccNum{n}%
\global\long\def\modN{\mathrm{mod\,}\ccNum}%
\newcommandx\kkCC[1][usedefault, addprefix=\global, 1=\ccIx]{\kk [#1]}%
\newcommandx\kkMCC[1][usedefault, addprefix=\global, 1=\ccIx]{\kkM [#1]}%

\newcommandx\wwdj[1][usedefault, addprefix=\global, 1=\dil]{r_{j}(#1)}%
\newcommandx\wwdjM[1][usedefault, addprefix=\global, 1=\dil]{r_{j}^{-}(#1)}%
\newcommandx\csr[1][usedefault, addprefix=\global, 1=i]{k_{#1}}%
\global\long\def\csrIxx{j}%
\global\long\def\CC{\gamma}%
\global\long\def\mm{a}%
\global\long\def\modelN{N}%

\global\long\def\aaass{a^{*}}%
\global\long\def\xxssi{x_{i}^{*}}%
\global\long\def\epssi{\sigma_{i}^{*}}%
\global\long\def\epssione{\sigma_{1}^{*}}%

\global\long\def\GG#1{\Delta G_{#1}^{\circ}/RT}%
\global\long\def\GGfrac#1{\frac{\Delta G_{#1}^{\circ}}{RT}}%
\global\long\def\chemodil{\dil}%

\global\long\def\sigmaj{\sigma_{(j)}}%

\catcode`\_=11
\definearrow4{<U=>}{%
    \CF_arrowshiftnodes{#4}%
    \path[allow upside down](\CF_arrowstartnode)--(\CF_arrowendnode)%
            node[pos=0,yshift=1pt](\CF_arrowstartnode u0){}%
            node[pos=0,yshift=-1pt](\CF_arrowstartnode d0){}%
            node[pos=1,yshift=1pt](\CF_arrowstartnode u1){}%
            node[pos=1,yshift=-1pt](\CF_arrowstartnode d1){};%
    \begingroup%
        \pgfarrowharpoontrue%
        \expandafter\draw\expandafter[\CF_arrowcurrentstyle](\CF_arrowstartnode u0)--(\CF_arrowstartnode u1)node[pos=0.4](Uarrowarctangent){};%
        \expandafter\draw\expandafter[\CF_arrowcurrentstyle](\CF_arrowstartnode d1)--(\CF_arrowstartnode d0);%
        \expandafter\draw\expandafter[\CF_arrowcurrentstyle](\CF_arrowstartnode u0)--(\CF_arrowstartnode u1)node[pos=0.8,yshift=-9.5pt](UarrowarctangentR){};%
    \endgroup%
    \expandafter\draw[semithick] \expandafter(Uarrowarctangent) arc (270:190:.333) node (Uarrowend) {};%
    \node[anchor=south,yshift=2pt] at (Uarrowend.north) {#1};
    \expandafter\draw[<-,>=stealth,semithick] \expandafter(UarrowarctangentR) arc (0:90:.333) node (Uarrowend) {};%
    \node[anchor=south,yshift=-22pt] at (Uarrowend.north) {#2};
}

\definearrow4{<V=>}{%
    \CF_arrowshiftnodes{#4}%
    \path[allow upside down](\CF_arrowstartnode)--(\CF_arrowendnode)%
            node[pos=0,yshift=1pt](\CF_arrowstartnode u0){}%
            node[pos=0,yshift=-1pt](\CF_arrowstartnode d0){}%
            node[pos=1,yshift=1pt](\CF_arrowstartnode u1){}%
            node[pos=1,yshift=-1pt](\CF_arrowstartnode d1){};%
    \begingroup%
        \pgfarrowharpoontrue%
        \expandafter\draw\expandafter[\CF_arrowcurrentstyle](\CF_arrowstartnode u0)--(\CF_arrowstartnode u1)node[pos=0.4](Uarrowarctangent){};%
        \expandafter\draw\expandafter[\CF_arrowcurrentstyle](\CF_arrowstartnode d1)--(\CF_arrowstartnode d0);%
        \expandafter\draw\expandafter[\CF_arrowcurrentstyle](\CF_arrowstartnode u0)--(\CF_arrowstartnode u1)node[pos=0.8,yshift=-9.5pt](UarrowarctangentR){};%
    \endgroup%
    \expandafter\draw[semithick] \expandafter(Uarrowarctangent) arc (270:190:.333) node (Uarrowend) {};%
    \node[anchor=south,yshift=2pt] at (Uarrowend.north) {#1};
    \expandafter\draw[<-,>=stealth,semithick] \expandafter(UarrowarctangentR) ; %arc (0:90:.333) node (Uarrowend) {};%
}

\ifrsb
 \newcommand\negpadA{\!\!\!}
 \newcommand\pospadA{\;\;\;\;\;\;}

 \newcommand\appdegradation{Section~A}
 \newcommand\appmultistep{Section~B}
 \newcommand\appchemostat{Section~C}

\else
 \newcommand\appdegradation{Section~\ref{app:degradation}}
 \newcommand\appmultistep{Section~\ref{app:multistep2}}
 \newcommand\appchemostat{Section~\ref{app:chemostat}}

 \def\thesection{\arabic{section}}
 \def\theequation{\arabic{section}.\arabic{equation}}
 \def\thesubsection{(\alph{subsection})}
 \makeatletter
 \def\@seccntformat#1{\csname the#1\endcsname\quad}
 \makeatother
 \counterwithin*{equation}{section}

 \newcommand\negpadA{ }
 \newcommand\pospadA{ }

\fi

\catcode`\_=8
\newcommand{\multistepScheme}[2]{
 \setchemfig{arrow coeff=0.6}
 \schemestart
 $#1$
 \arrow{<U=>[{\negpadA\footnotesize $\sum_i \alpha_{1,i} \MMi$}][{\negpadA\footnotesize $\sum_i \beta_{1,i} \MMi$}]}
 $Y_{1}$
 \arrow{<U=>[{\pospadA\footnotesize $\sum_i \alpha_{2,i} \MMi$}][{\pospadA\footnotesize $\sum_i \beta_{2,i} \MMi$}]}
 $\dots$
 \arrow{<U=>[$\cdots$][$\quad\quad\cdots$]}
 $Y_{m-1}$
 \arrow{<U=>[{\footnotesize $\sum_i \alpha_{m,i} \MMi$}][{\footnotesize $\sum_i \beta_{m,i} \MMi$}]}
 $ #2$
 \schemestop
}

\begin{document}

\title{Thermodynamics of Darwinian selection in molecular replicators}

\newcommand{\addressupf}{ICREA-Complex Systems Lab, Universitat Pompeu Fabra, 08003 Barcelona,
Spain}
\newcommand{\addressubi}{Universal Biology Institute, The University of Tokyo, 7-3-1 Hongo,
Bunkyo-ku, Tokyo 113-0033, Japan}

\ifrsb
    \author{Artemy Kolchinsky$^{1}$}
    %%%%%%%%% Insert author address here
    \address{$^{1}$\addressupf\\
    $^{2}$\addressubi}

    %%%% Subject entries to be placed here %%%%
    \subject{thermodynamics, evolution, origin of life}

    %%%% Insert corresponding author and its email address}
    \corres{Artemy Kolchinsky\\
    \email{artemyk@gmail.com}}

\else 
    \author{Artemy Kolchinsky}
    \email{artemyk@gmail.com}
    \affiliation{\addressupf}
    \affiliation{\addressubi\looseness=-1}
\fi

\begin{abstract}
We consider the relationship between thermodynamics, fitness, and
Darwinian selection in autocatalytic molecular replicators. We uncover
a thermodynamic bound that relates fitness, replication rate, and
thermodynamic affinity of replication. This bound applies to a
broad range of systems, including elementary and non-elementary autocatalytic
reactions, polymer-based replicators, and certain kinds of autocatalytic
sets. In addition, we show that the critical selection coefficient (the minimal fitness difference visible to selection) is bounded by a simple function of the affinity.  Our results imply
fundamental thermodynamic bounds on selection strength in molecular
evolution, complementary to other bounds that arise from finite population
sizes and error thresholds. These bounds may be relevant for understanding
thermodynamic constraints faced by early replicators at the origin
of life. We illustrate our approach on several examples, including
a classic model of replicators in a chemostat.
\end{abstract}  

\ifrsb
    %%%% Keyword entries to be placed here %%%%

    \rsbreak

\else
    \maketitle
\fi

\section{Introduction}

Recent work has uncovered fundamental bounds on the thermodynamic
costs of various biomolecular processes, including chemical sensing~\cite{mehta2012energetic,barato_efficiency_2014,govern2014energy},
copying of polymer-stored information~\cite{andrieuxNonequilibriumGenerationInformation2008,ouldridge2017fundamental,poultonNonequilibriumCorrelationsMinimal2019,sartoriKineticEnergeticDiscrimination2013},
and growth and replication~\cite{kondo_growth_2011,england2013statistical,himeoka_entropy_2014,virgo2016complex,saakianNonlinearStochasticDynamics2016,bishopStochasticBistabilityBifurcation2010,pinero_nonequilibrium_2018,corominas2019thermodynamics}.
These results are derived from general principles of nonequilibrium
thermodynamics --- such as flux-force relations and fluctuation theorems~\cite{beard2007relationship,jarzynski_equalities_2011,seifert2012stochastic,kondepudiModernThermodynamicsHeat2015}
--- that relate the dynamical and thermodynamic properties
of nonequilibrium processes. Due to their generality, these bounds
shed light on universal thermodynamic constraints on lifelike systems,
including modern and protobiological organisms, synthetic life, and
even possible non-terrestrial lifeforms.

One of the most important properties of living systems is that they
undergo Darwinian selection. Generally speaking, Darwinian selection
refers to a process in which high-fitness replicators reliably 
outcompete low-fitness replicators. Darwinian selection can
be exhibited by chemical systems, such as individual replicating molecules
or networks of molecules~\cite{eigenSelforganizationMatterEvolution1971,schusterDynamicsEvolutionaryOptimization1985,leeAutocatalyticNetworksTransition1997,feistelPhysicsSelfOrganization2011,vasasEvolutionGenes2012a,takeuchiEvolutionaryDynamicsRNAlike2012,vaidyaSpontaneousNetworkFormation2012,szilagyi2017ecology,peng2020ecological,ameta2021self,mizuuchiEvolutionaryTransitionSingle2022,ametaDarwinianPropertiesTheir2021}.
Synthesizing such systems has been a major focus of research on the
origin of life, given that the emergence of Darwinian selection is
considered to be a crucial point in the transition from nonliving to living
matter~\cite{bernstein1983darwinian,yarusGettingRNAWorld2011,vasasEvolutionGenes2012a,ameta2021self,jeancolasThresholdsOriginLife2020}. 

Here, we consider the relationship between thermodynamics and Darwinian
selection in minimal chemical systems. This relationship may be particularly
relevant for understanding thermodynamic constraints on the origin
of life~\cite{pascal2013towards,jeancolasThresholdsOriginLife2020,dangerConditionsMimickingNatural2020}. 
We consider a reactor containing autocatalytic
replicators that copy themselves either via elementary
reactions, or via more complex multi-step mechanisms.
We also consider certain types of collectively autocatalytic sets,
where replication involves a cycle of cross-catalytic reactions.
Our setup includes many types of molecular replicators previously
considered in the literature, including self-complementary and complementary
templates, polymer-based replicators, and autocatalytic small molecules~\cite{patzke2007self,bissetteMechanismsAutocatalysis2013a,lincolnSelfsustainedReplicationRNA2009,leeSelfreplicatingPeptide1996}.
It also encompasses several classic models of molecular replicators,
including the chemostat model~\cite{schusterDynamicsEvolutionaryOptimization1985,feistelPhysicsSelfOrganization2011}
and Eigen's quasispecies model~\cite{eigenSelforganizationMatterEvolution1971}.

Each replicator is associated with three quantities. The first is
the \emph{affinity}  $\ep$ of the replication reaction, the thermodynamic driving force of the reaction. 
The second is
the \emph{replication rate} $\rr$, the number of copies a replicator makes
per unit time under actual conditions. Lastly, in Section~\ref{subsec:Fitness}, we define the \emph{fitness}
$f$ of a replicator as the maximum achievable replication rate, reached in the
limit of small concentrations. We show that
fitness determines a replicator's ability to invade a given population~\cite{metzHowShouldWe1992} and to survive a high dilution rate. 
The introduction of an operational definition of fitness for molecular
replicators is an important contribution of our work.

In Section~\ref{sec:darw}, we derive a thermodynamic bound that relates affinity $\ep$,  replication rate $\rr$, and fitness $f$ 
as
\begin{equation}
\ep\ge-\ln\left(1-\frac{\rr}{f}\right).\label{eq:introbnd}
\end{equation}
As we discuss below, affinity is a fundamental thermodynamic cost that captures %represents the driving force of the chemical reaction, or equivalently 
the dissipated Gibbs free energy (entropy production) in a single replication event. Eq.~\eqref{eq:introbnd} implies that a minimum affinity is required to sustain a given replication rate, and that this minimum increases as the replication
rate approaches its maximum possible value, the fitness. Conversely, Eq.~\eqref{eq:introbnd} implies that for a given affinity, there is a fundamental limit on how closely the replication
rate can approach its maximum value.

We also derive a thermodynamic bound on the strength of Darwinian
selection. Observe that a higher fitness replicator is not always
able to outcompete a lower fitness one, as this depends on the fitness difference
as well as various environmental and demographic factors~\cite{gillespie2004population}. Selection strength 
can be quantified in terms of the smallest fitness difference that
can affect evolutionary outcomes in a given population and environment.
This is the so-called \emph{critical selection coefficient} $s$,
the ``resolution limit'' below which fitness
differences are indiscernible. 

To use a well-known example, selection strength in finite populations is limited by the stochasticity of sampling, and a
fitter mutant will fixate with high probability only
when $s\gg1/N_{e}$, where $N_{e}$ is the effective population size~\cite{ewensMathematicalPopulationGenetics2004}. As another example,
Eigen's ``error threshold'' implies that selection strength is limited by the mutation rate $\mu$, so
that a fitter mutant can dominate the population only when $s>\mu$~\cite{eigenSelforganizationMatterEvolution1971,smithMajorTransitionsEvolution}. % \cite[Eq.~II-45,][]

In Section~\ref{sec:darw}, we use the inequality~\eqref{eq:introbnd} to derive a thermodynamic
bound on the critical selection coefficient. We suppose that selection is sufficiently strong so that a replicator with fitness
$f$ is present in the steady state of a flow reactor, while another
replicator type with fitness $f'<f$ is driven to extinction. We show
that $s=1-f'/f$, the selection coefficient between the two replicators,
must obey 
\begin{equation}
s\ge e^{-\epss},\label{eq:introres}
\end{equation}
where $\epss$ is the affinity of the fitter replicator in steady state. 
This bound on the strength of selection
applies even in the case of infinite populations and error-free replicators.
It implies that selecting for a relative fitness difference of $s$
must dissipate more than $-\ln s$ of free energy,  a quantity that 
 diverges in the limit of vanishing fitness differences, $s\to0$. We illustrate
this result using a classic model of replicators in a chemostat in
Section~\ref{sec:example}.

In Section~\ref{sec:crosscatalytic}, we extend our results to autocatalytic
sets, where replication involves a cycle of cross-catalytic reactions.
To do so, we first generalize our notion of fitness to autocatalytic sets and
then derive generalized versions of the inequalities~\eqref{eq:introbnd} and~\eqref{eq:introres}.
In these generalizations, $\ep$ refers to the affinity of the average cross-catalytic reaction in the cycle.

\begin{figure}
\ifrsb
    \newcommand{\figAwidth}{160pt}
    \newcommand{\figAboxwidth}{0.35\textwidth}
\else
    \newcommand{\figAwidth}{0.7\columnwidth}
    \newcommand{\figAboxwidth}{0.45\textwidth}
\fi
\begin{center}
\begin{minipage}[c]{\figAboxwidth}
\includegraphics[width=\figAwidth]{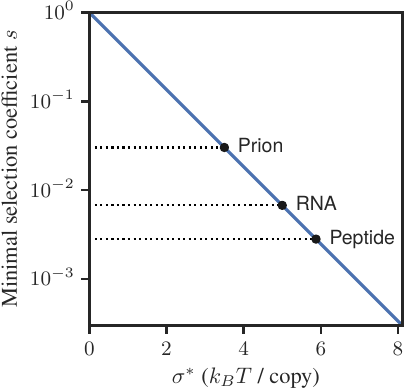}
\ifrsb
\else
\vspace{15pt}
\fi
\end{minipage}
%\hfill
\begin{minipage}[c]{0.45\textwidth}
\centering
\begin{tblr}{colspec={|Q[l]|Q[l,3cm]|Q[l]|},rowspec={|Q[m]||Q[m]|Q[m]|Q[m]|}}
 \textbf{Replicator} & \textbf{Concentrations} & $-\Delta G^\circ$\\ 
 {Prion~\cite{prusiner1991molecular}} & Equal native and misfolded, pH 3.6
 & {Table~1~\cite{baskakovFoldingPrionProtein2001}} \\  
 RNA~\cite{lincolnSelfsustainedReplicationRNA2009} & \raggedright  \emph{In vivo}~\cite{julicherMotionRNAPolymerase1998,erieSingleNucleotideAdditionCycle1992}
 & \emph{In vivo}~\cite{julicherMotionRNAPolymerase1998,erieSingleNucleotideAdditionCycle1992} \\
 Peptide~\cite{leeSelfreplicatingPeptide1996} & \raggedright Substrate $90\,\mu\mathrm{M}$, replicator $5\,\mu\mathrm{M}$~\cite{leeSelfreplicatingPeptide1996}   
 & {Figure~4~\cite{wangTheoreticalAnalysisDetailed2011}}
 \end{tblr} 
\end{minipage}
\end{center}
\caption{\label{fig:empirical}Illustration of our thermodynamic bound~\eqref{eq:introres} 
for three real-world molecular replicators: a prion~\cite{baskakovFoldingPrionProtein2001}, an RNA molecule that copies
itself via a single ligation~\cite{lincolnSelfsustainedReplicationRNA2009},
and a peptide that copies itself via ``native chemical ligation''~\cite{leeSelfreplicatingPeptide1996,dawsonSynthesisProteinsNative1994}. 
Affinities were computed using Eq.~\eqref{eq:EPdef2} from the concentrations and standard Gibbs energies $-\Delta G^\circ$ listed in the table (at room temperature). 
Note
that there is some debate whether prion replication is first-order,
like the replicators considered in this paper, or instead involves
higher-order cooperative interactions~\cite{eigenPrionicsKineticBasis1996,laurentAutocatalyticProcessesCooperative1997a,laurentPrionDiseasesProtein1996,fematMechanismsPrionDisease2011}. 
}
\end{figure}

In principle, our results can also be applied to various real-world
molecular replicators, suggesting a route for experimental validation.
In Figure~\ref{fig:empirical}, we use published thermodynamic data to illustrate
the bound~\eqref{eq:introres} on three real-world replicators. The
first is a prion at low pH~\cite{prusiner1991molecular}, where
$\epss\approx3.5$ (assuming equal concentrations of native and misfolded form)~\cite{baskakovFoldingPrionProtein2001}. 
The second is an RNA molecule that copies itself using a single RNA ligation~\cite{lincolnSelfsustainedReplicationRNA2009}, with $\epss\approx5$ under \emph{in vivo} conditions~\cite{julicherMotionRNAPolymerase1998,erieSingleNucleotideAdditionCycle1992}. The third is a peptide that copies itself using ``native chemical ligation''~\cite{leeSelfreplicatingPeptide1996,wangTheoreticalAnalysisDetailed2011}, 
with $\epss\approx5.9$ at published concentrations~\cite{leeSelfreplicatingPeptide1996}. 
%The second system is a peptide that copies itself using
%``native chemical ligation''~\cite{leeSelfreplicatingPeptide1996,wangTheoreticalAnalysisDetailed2011},
%with $\epss\approx5.9$ at published concentrations~\cite{leeSelfreplicatingPeptide1996}.
%The third is an RNA molecule that copies itself using
%a single RNA ligation~\cite{lincolnSelfsustainedReplicationRNA2009},
%with $\epss\approx5$ under \emph{in vivo} conditions~\cite{julicherMotionRNAPolymerase1998,erieSingleNucleotideAdditionCycle1992}. 
For example, our result predicts that for the RNA replicator, selection
can only discern relative fitness differences of $s\ge e^{-5}\approx0.6\%$.

\section{Setup}

\label{sec:setup}

We consider a chemical reactor at constant temperature and pressure.
The reactor contains an ideal well-mixed solution of replicators
$X,X^{\prime},\dots$ and other chemical species $A_{1},A_{2},\dots$
that may serve as substrates and side products. We study this system
in terms of deterministic concentrations, assuming that molecular
counts are sufficiently large so that stochastic fluctuations can
be ignored. We use $x$ to indicate the concentration of replicator
$X$ and $\aa=(a_{1},a_{2},\dots)$ to indicate the concentrations
of substrates/side products $(A_{1},A_{2},\dots)$.

Each replicator $X$ undergoes a reversible autocatalytic reaction
of the form
\begin{equation}
X+\sum_{i=1}\alpha_{i}\MMi\rightleftharpoons X+X+\sum_{i=1}\beta_{i}\MMi,\label{eq:autocatalysis}
\end{equation}
where $\alpha_{i}$ and $\beta_{i}$ are stoichiometric coefficients
of substrate and side products $A_{i}$. A simple example of Eq.~\eqref{eq:autocatalysis}
is autocatalysis from a single substrate, $X+\MM\rightleftharpoons X+X$,
but many other schemes are also possible. The autocatalytic reaction
may be elementary, or it may proceed via multiple steps. Different
types of replicators will generally have different stoichiometric
coefficients $\alpha_{i},\beta_{i}$ as well as other thermodynamic
and kinetic parameters (discussed below). 

We use $\mathcal{J}$ to indicate the net flux across the autocatalytic
reaction in Eq.~\eqref{eq:autocatalysis}. We define the replication rate
$\rr$ of replicator $X$ as the net flux per replicator,
\begin{equation}
\rr:=\frac{\mathcal{J}}{x}.\label{eq:ggdef}
\end{equation}
The replicators may also flow out of the volume with dilution rate
$\dil\ge0$. Accounting for both replication and dilution, the concentration
of replicator $X$ changes as
\begin{equation}
\dot{x}=\mathcal{J}-\dil x=(\rr-\dil)x.\label{eq:dx}
\end{equation}
We usually leave dependence on time $t$ implicit in our notation.
We will use that in steady state, any non-extinct replicator ($x>0$)
must have $\rr=\dil$, meaning that dilution and replication balance.

In writing Eq.~\eqref{eq:dx}, we assume that different
replicators do not interact directly by consuming each other as substrates
or producing each other as side products, although they may interact
indirectly via shared substrates/side products $A_{i}$. For simplicity,
we also ignore the spontaneous degradation of replicators. However,
in \appdegradation{} of the Supplementary Material (SM), we show that our results still hold in
the presence of degradation reactions.

Eq.~\eqref{eq:dx} also assumes that the rate of spontaneous (i.e., non-autocatalytic) formation of the replicator is negligible. This assumption plays an important role in our analyses below, since replicators that can form spontaneously do not exhibit first-order growth, nor do they go completely extinct in steady-state even at  large dilution rates. An interesting direction for future work would extend our analysis to replicators with non-negligible rates of spontaneous formation.

As will be noted below, some of our results hold for closed reactors as well as open
reactors that exchange matter with their environment~\cite{avanzini2022thermodynamics}.
However, our thermodynamic bound on selection~\eqref{eq:introres}
applies specifically to a flow reactor in steady state. Different
types of flow reactors may be considered. One example is the continuous
stirred-tank reactor (CSTR) where the dilution rate and inflow rates
are constant, which is often used in chemical~\cite{filisettiStochasticModelEmergence2011a,semenovAutocatalyticBistableOscillatory2016}
and biological experiments~\cite{dykhuizen1983selection,groverResourceCompetition1997,hoskisson2005continuous},
and which can also arise naturally (e.g., in a pond fed by a nutrient-rich
stream). To avoid confusion, we note that the CSTR is called a \emph{chemostat}
in the biological literature~\cite{harmandChemostat2017,smithTheoryChemostatDynamics1995},
although in nonequilibrium thermodynamics, the term \emph{chemostat}
sometimes refers instead to an external chemical reservoir~\cite{polettini2014irreversible}. Another possible flow
reactor is one where the rates of dilution and inflow can vary as
a function of the chemical concentrations. An example is provided
by Eigen's quasispecies model, where the dilution rate is adjusted
to maintain the total concentration of replicators constant~\cite{eigenSelforganizationMatterEvolution1971}. 

Our main thermodynamic quantity of interest is the 
affinity $\ep$ of the replication reaction~\eqref{eq:autocatalysis}. A reaction proceeds in the forward direction if and only if the affinity is positive, so affinity can be understood as the driving force of the chemical reaction. Equivalently, the affinity is proportional to the Gibbs energy of reaction, the free energy dissipated in a single reaction event. This dissipated free energy
is a fundamental thermodynamic cost that 
represents
lost work potential: a reaction with affinity $\ep$
can be coupled to a thermodynamically disfavored ``uphill''
reaction and thereby perform up to $\ep$ of chemical work per reaction event.

For an ideal solution, we write the affinity in dimensionless units as 
\begin{equation}
\ep=-\ln x +\sum_{i=1}(\alpha_{i}-\beta_{i})\ln a_{i}-\Delta G^{\circ}/RT,\label{eq:EPdef2}
\end{equation}
where $R$ is the gas constant, $T$ is the temperature, and $-\Delta G^{\circ}$
is the standard Gibbs energy (in units of joules/mole),
\begin{equation}
-\Delta G^{\circ}=RT\Big[\ln x^{\text{eq}}-\sum_{i=1}(\alpha_{i}-\beta_{i})\ln a_{i}^{\text{eq}}\Big].\label{eq:stdg}
\end{equation}
Here $x^{\mathrm{eq}}$ and $a_{i}^{\mathrm{eq}}$ are equilibrium
concentrations of $X$ and $A_{i}$, as would be reached if the reactor
was closed to exchange of matter and allowed to relax completely. 
As usual, logarithms of concentrations, as in Eqs.~\eqref{eq:EPdef2} and \eqref{eq:stdg}, should be considered dimensionless after dividing by the standard concentration (e.g., dividing by $c^\circ = 1\;\textrm{M}$ if $x$ and $a_i$ are expressed in units of molar concentration).

%It is clear that $\ep$ vanishes in equilibrium, when there is no remaining potential for doing work.

We emphasize again that we express the affinity $\ep$ in dimensionless units. 
It can be understood in terms of energy units as the number of $k_B T$ dissipated when making a single replicator copy, where 
$k_B$ is Boltzmann's constant and $k_B T$ is the typical energy (in joules) of a thermal fluctuation at temperature $T$. 
%It can be converted to joules per copy as $k_{B}T\sigma$, where
%$k_{B}$ is Boltzmann's constant, or to joules per mole as $RT\sigma$.
In the chemistry literature, $\ep$ is
often written in units of joules/mole as $-\Delta G=RT \sigma$.

\section{Elementary and non-elementary replicators}

\label{sec:nelem}

\subsection{Elementary replicators}

\label{subsec:Elementary-replicators}

An \emph{elementary replicator} refers to the case where Eq.~\eqref{eq:autocatalysis}
is an elementary reaction. Then, the net flux across the reaction
has the mass-action form~\cite{kondepudiModernThermodynamicsHeat2015}
\begin{equation}
\mathcal{J}=\ww x-\wwM x^{2},\label{eq:jelem}
\end{equation}
with forward and backward rate constants 
\begin{equation}
\ww=\kappa\prod_{i}a_{i}^{\alpha_{i}},\qquad\wwM=\kappa e^{\Delta G^{\circ}/RT}\prod_{i}a_{i}^{\beta_{i}},\label{eq:massacction}
\end{equation}
and $\kappa$ is a baseline rate constant. Note that $\ww$ and $\wwM$ may
depend on the concentrations of substrates/side products $\aa$ (technically, they are ``pseudo rate constants''). We leave this
dependence implicit in our notation.

For elementary replicators, the log-ratio of the forward and backward fluxes in Eq.~\eqref{eq:jelem}
equals the affinity of the replication reaction,
\begin{equation}
\ep=\ln\frac{\ww\xx}{\wwM\xx^{2}}=\ln\frac{\ww}{\wwM x}\,,\label{eq:fluxforce3}
\end{equation}
as can be verified by comparing Eqs.~\eqref{eq:massacction} and \eqref{eq:EPdef2}.
Eq.~\eqref{eq:fluxforce3} is an instance of a general result, called the
\emph{flux-force relation} or \emph{local detailed balance} in the
literature~\cite{wachtelThermodynamicallyConsistentCoarse2018,raoNonequilibriumThermodynamicsChemical2016},
which relates the affinity of an elementary reaction with
the forward and reverse fluxes~\cite{kondepudiModernThermodynamicsHeat2015}.
The flux-force relation is one of the most important results in nonequilibrium
thermodynamics, since it connects the kinetic properties of a chemical
reaction with its thermodynamic properties.

In fact, our results do not  require that the rate constants $\ww$
and $\wwM$ have the mass action form of Eq.~\eqref{eq:massacction}, only
that Eq.~\eqref{eq:jelem} and Eq.~\eqref{eq:fluxforce3} hold. In principle, this could be
used to study certain non-ideal solutions that exhibit interactions
between $A_{i}$~\cite{avanzini2021nonequilibrium}.

\subsection{Non-elementary replicators}

\label{subsec:nonelemkinetic}

Most real-world autocatalytic replicators cannot be treated as elementary
reactions. For this reason, we also consider the case where Eq.~\eqref{eq:autocatalysis}
represents a reaction mechanism that proceeds via a sequence of $\msNum$
steps,
\begin{equation}
\multistepScheme{X}{X+X}.\label{eq:multistep}
\end{equation}
Each $Y_{k}$ is an intermediate chemical species, and each intermediate
step is an elementary reversible reaction with mass-action kinetics
that may involve substrates/side products $\MMi$ with stoichiometric
coefficients $\alpha_{\msIx,i}$ and $\beta_{\msIx,i}$. The stoichiometry
of the overall reaction is $\alpha_{i}=\sum_{\msIx}\alpha_{\msIx,i}$
and $\beta_{i}=\sum_{\msIx}\beta_{\msIx,i}$. We assume that intermediate
species are not shared between different types of replicators. For
simplicity, we also assume that degradation reactions are negligible,
although our results generalize to the presence of degradation as
shown in \appdegradation{} in the SM.

\begin{figure}
\ifrsb
    \newcommand{\figBwidth}{300pt}
\else
    \newcommand{\figBwidth}{1\columnwidth}
\fi
\begin{center}
\includegraphics[width=\figBwidth]{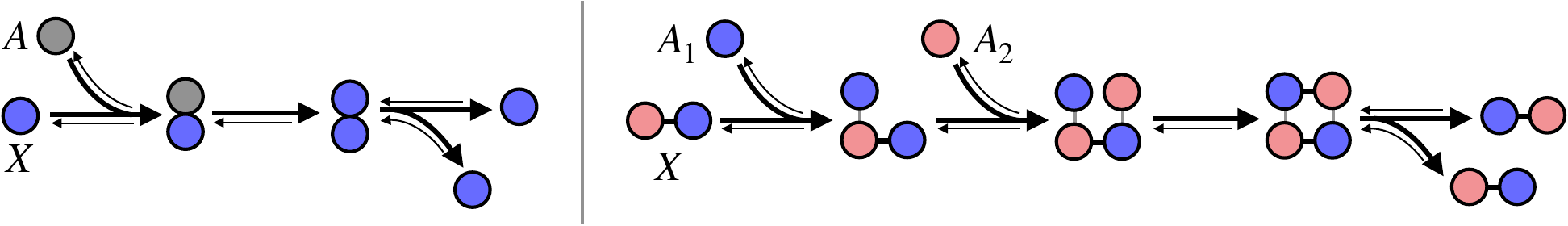}
\end{center}
\caption{\label{fig:complex}Examples of non-elementary autocatalytic replication
mechanisms. \emph{Left}: autocatalysis with binding, conversion, and
unbinding steps. \emph{Right}: templated replication of a self-complementary
polymer (shown here using a dimer).}
\end{figure}

We term this kind of reaction mechanism a \emph{non-elementary replicator},
although it is also called an ``autocatalytic cycle'' in the literature~\cite{kingAutocatalysis1978,virgo2016complex,hordijkAutocatalyticConfusionClarified2017}.
A simple example of a non-elementary replicator is a three-step mechanism
with binding, conversion, and unbinding steps, see Figure~\ref{fig:complex}~(left).
Another example is the step-by-step replication of a self-complementary
dimer, illustrated in Figure~\ref{fig:complex}~(right), which has been studied in numerous
origin-of-life experiments~\cite{von1986self,patzke2007self,bissetteMechanismsAutocatalysis2013a}.
Yet other examples include the formose cycle~\cite{bissetteMechanismsAutocatalysis2013a}
and the autophosphorylation of protein kinase~\cite{wangAutophosphorylationKineticsProtein2002,bishopStochasticBistabilityBifurcation2010}.

In this subsection, we show that under reasonable assumptions, the production rate and affinity of non-elementary replicators can be expressed in a simple form, somewhat analogous
to Eq.~\eqref{eq:massacction} and Eq.~\eqref{eq:fluxforce3}. 
Specifically, we will show that the production of a non-elementary replicator can be written in a mass-action-like form,
\begin{equation}
\mathcal{J}=\wwd[\rr]\xx-\wwdM[\rr]\xx^{2},\label{eq:jelem-1}
\end{equation}
with effective rate constants $\wwd[\rr]$ and $\wwdM[\rr]$ that depend explicitly on the replication rate $\rho$ and implicitly on the dilution rate and the rate constants of intermediate reactions. We will also show that the affinity of a non-elementary replication reaction can be expressed in a ``flux-force-like'' form,
\begin{equation}
\ep=\ln\frac{x\wwd[0]}{\xx^{2}\wwdM[0]}=\ln\frac{\wwd[0]}{\xx\wwdM[0]}.\label{eq:fluxforcemech-1}
\end{equation}
Given Eq.~\eqref{eq:jelem-1}, the numerator and denominator
can be interpreted as the forward and backward fluxes in the case
of $\rr \to 0$, the limit when dilution is much slower than internal
reactions. Ref.~\cite{wachtelThermodynamicallyConsistentCoarse2018}
previously derived this flux-force-like relation for the special case of steady-state conditions. 
%and reaction mechanisms such as Eq.~\eqref{eq:multistep}, which involve
%a single reaction cycle.

The rest of this subsection is devoted to the   derivation of Eqs.~\eqref{eq:jelem-1} and \eqref{eq:fluxforcemech-1}. Readers not interested in this derivation may skip it without affecting their comprehension of the main message of the paper.

Observe that each intermediate reaction $k\in\{1,\dots,m\}$ has net flux
\begin{align}
\JJ[\msIx] & =\vv y_{\msIx-1}-\vvM y_{\msIx}.\label{eq:yflux}
\end{align}
Here $y_{k}$ is the concentration of intermediate species $Y_{\msIx}$
for $k\in\{1,\dots,m-1\}$, and we use the convention $y_{0}=x,y_{m}=x^{2}$.
The terms $\rrArg{\msIx}$ and $\rrMArg{\msIx}$ refer to forward
and backward (pseudo) rates constants of intermediate steps, typically
defined in a manner analogous to Eq.~\eqref{eq:massacction}. We note that
$\rrArg{\msIx}$ and $\rrMArg{\msIx}$ can depend on the concentrations
$\aa$, although we leave this implicit in our notation.

The rate of autocatalytic production of replicator $X$ is
\begin{equation}
\mathcal{J}=2\JJ[\msNum]-\JJ[1],\label{eq:netflux}
\end{equation}
since two copies are produced by the last step and one consumed by
the first. The production of intermediate species $Y_{k}$ ($k\in\{1,\dots,m-1\}$)
due to the mechanism is $\JJ[\msIx]-\JJ[\msIx+1]$, since one $Y_{k}$
is produced by intermediate step $k$ and one is consumed by intermediate
step $k+1$. The rate of change of the concentration of the intermediate
species, also accounting for dilution with rate $\dil\ge0$, is 
\begin{equation}
\dot{y}_{k}=\JJ[\msIx]-\JJ[\msIx+1]-\dil y_{k}.\label{eq:derivs}
\end{equation}

To derive Eq.~\eqref{eq:jelem-1}, %It will be useful to express the overall flux $\mathcal{J}$ in a
% mass-action-like form of Eq.~\eqref{eq:jelem}. To do so, 
we introduce the
assumption that the relative concentrations of intermediate species
$Y_{k}$ and replicator $X$ is approximately constant, 
\begin{equation}
\frac{d}{dt}\frac{y_{k}}{x}=\frac{x\dot{y}_{k}-\dot{x}y_{k}}{x^{2}}\approx0.\label{eq:qa}
\end{equation}
This assumption always holds in steady state ($\dot{y}_{k}=\dot{x}=0$).
It also holds under a separation of timescales where relative
concentrations $y_{k}/x$ relax to steady-state values much faster than 
these steady-state values change (e.g., as a result of changing dilution rate, concentrations
$\vec{a}$, etc.). This separation of timescales is valid during
the exponential growth phase of an initially rare replicator, when
the absolute concentrations of $x$ and $y_{k}$ are small and
have a minimal effect on other variables.

Our assumption~\eqref{eq:qa} of stationary \emph{relative} concentrations is somewhat different from 
the quasi-steady-state (QSS) approximation, as often employed in biochemistry~\cite{segel1989quasi} and recently considered in nonequilibrium thermodynamics~\cite{wachtelThermodynamicallyConsistentCoarse2018,avanzini2023circuit}. QSS assumes that the \emph{absolute} concentrations
of intermediate species $y_k$ are approximately stationary, compared to
the rate of change of the replicator concentration $x$. This assumption  
can be violated in growing autocatalytic replicators, even if the \emph{relative} concentrations are nearly stationary (e.g., when the intermediate species and replicator grow quickly but at the same rate).

Plugging $\dot{x}=(\rr-\dil)x$ and $\dot{y}_{k}=\JJ[\msIx]-\JJ[\msIx+1]-\dil y_{k}$
into Eq.~\eqref{eq:qa} and simplifying gives
\begin{equation}
\JJ[\msIx]-\JJ[\msIx+1]=\rr y_{k}.\label{eq:QSS2}
\end{equation}
Using Eq.~\eqref{eq:yflux}, we express the intermediate concentrations
$\yy=(y_{1},\dots,y_{m-1})$ in terms of a linear system
\begin{equation}
[(M+\rr I)\yy]_{k}=\delta_{k,1}\rrArg 1x+\delta_{k,m-1}\rrMArg{\msNum}x^{2},\label{eq:yvec0}
\end{equation}
where $\Amatrix\in\mathbb{R}^{(m-1)\times(m-1)}$ is a matrix of intermediate
rate constants,
\begin{equation}
\Amatrix\!=\!\left[\begin{array}{ccccc}
\rrMArg 1\!\!+\!\rrArg 2 & -\rrMArg 2 & 0 & 0 & \dots\\
-\rrArg 2 & \rrMArg 2\!\!+\!\rrArg 3 & -\rrMArg 3 & 0 & \dots\\
0 & \dots & \dots & \dots & -\rrMArg{\msNum-1}\\
\dots & 0 & 0 & -\rrArg{\msNum-1} & \rrMArg{\msNum-1}\!\!+\!\rrArg{\msNum}
\end{array}\right]\label{eq:mmatrix-2}
\end{equation}
For any $\rr\ge0$, $\Amatrix+\rr I$ is an ``M-matrix'', so it is invertible and all entries of its inverse are nonnegative~\cite{shivakumar1996two,meyer1978singular}.

Finally, we solve Eq.~\eqref{eq:yvec0} for $\yy$ and combine with Eqs.~\eqref{eq:netflux} and \eqref{eq:yflux}
to write the production rate in the form of Eq.~\eqref{eq:jelem-1}, $\mathcal{J}=\wwd[\rr]\xx-\wwdM[\rr]\xx^{2}$. The effective rate constants are
\begin{align}
\begin{aligned}\wwd[\rr] & :=\rrArg 1(\rrMArg 1G_{11}+2\rrArg{\msNum}G_{m-1,1}-1)\\
\wwdM[\rr] & :=\rrMArg{\msNum}(2-2\rrArg{\msNum}G_{m-1,m-1}-\rrMArg 1G_{1,m-1})\,,
\end{aligned}
\label{eq:effphi-1}
\end{align}
where for convenience we defined the matrix
\begin{align}
G:=(\Amatrix+\rr I)^{-1}.
\end{align}
These effective rate constants
can depend on concentrations $\aa$ (since $\rrArg k$ and $\rrMArg k$
depend on them), although we leave this dependence implicit in our
notation.

Below we will use that the first and second derivatives of the effective
rate constants obey
\begin{equation}
\begin{aligned}\partial_{\rr}\wwd[\rr] & \le0,\pospadA\;\partial_{\rr}^{2}\wwd[\rr]\ge0\\
\partial_{\rr}\wwdM[\rr] & \ge0,\;\pospadA\partial_{\rr}^{2}\wwdM[\rr]\le0
\end{aligned}
\label{eq:derivsR}
\end{equation}
This follows from Eq.~\eqref{eq:effphi-1} by using matrix calculus and
the fact that all entries of $G$ are nonnegative.

For $\rr=0$, the effective rate constants can be expressed
in closed form,
\begin{align}
\wwd[0] & =\Bigg[\sum_{\msIx=1}^{\msNum}\frac{\prod_{\msIxx=1}^{\msIx-1}\rrMArg{\msIxx}}{\prod_{\msIxx=1}^{\msIx}\rrArg{\msIxx}}\Bigg]^{-1}\;\quad\pospadA\wwdM[0]=\wwd[0]\prod_{\msIx=1}^{\msNum}\frac{\rrMArg{\msIx}}{\rrArg{\msIx}}\label{eq:r0}
\end{align}
as shown in \appmultistep{} in the SM. The $\rr=0$ regime corresponds
to the limit where the replication rate $\rr$ is much slower than
the rate of internal reactions, so that $\rr$ can be neglected when
solving the linear system~\eqref{eq:yvec0}. 

The effective reverse rate constant $\wwdM[\rr]$ is always nonnegative, since $\wwdM[0]\ge 0$ and $\partial_\rr \wwdM[\rr]\ge 0$ from Eq.~\eqref{eq:derivsR}-\eqref{eq:r0}. On the other hand, formally the effective forward rate constant $\wwd[\rr]$  becomes negative for sufficiently large $\rr$, as can be seen from Eq.~\eqref{eq:effphi-1}  and $G\to 0$ as $\rho \to \infty$. However, in the following, we only consider the physically-meaning range of $\rr$ for which $\wwd[\rr]$ is nonnegative,  as discussed below in our definition of fitness~\eqref{eq:fitdef}.

\label{subsec:nonelemthermo}

We finish by discussing the thermodynamics of non-elementary replicators.
The affinity of intermediate reaction $k$ in Eq.~\eqref{eq:multistep}
is
\begin{equation}
\epK=\ln\frac{y_{\msIx-1}}{y_{\msIx}}+\sum_{i=1}(\alpha_{k,i}-\beta_{k,i})\ln a_{i}-\Delta G_{k}^{\circ}/RT,\label{eq:epk}
\end{equation}
where $y_{0}=x,y_{\msNum}=x^{2}$ (as above). Since intermediate
reactions are elementary, the forward and backward fluxes in Eq.~\eqref{eq:yflux}
obey the flux-force relation,
\begin{equation}
\epK=\ln\frac{\vv y_{\msIx-1}}{\vvM y_{\msIx}},\label{eq:epk-1}
\end{equation}
as can be shown explicitly when the rate constants $\vv$ and $\vvM$
have mass-action kinetics similar to Eq.~\eqref{eq:massacction}. 
The affinity of the overall replication mechanism is the
sum of the affinities of the individual steps~\cite{kondepudiModernThermodynamicsHeat2015},
\begin{equation}
\ep=\sum_{k=1}^{m}\epK=\ln\frac{1}{x}+\sum_{k=1}^{m}\ln\frac{\vv}{\vvM}.\label{eq:epSum}
\end{equation}
We arrive at Eq.~\eqref{eq:fluxforcemech-1} by combining Eqs.~\eqref{eq:r0} and \eqref{eq:epSum}.

Observe that an elementary replicator is a special case of a non-elementary replicator
with a single reaction ($m=1$) and no intermediate species. In this
case, the effective rate constants in Eq.~\eqref{eq:effphi-1} lose dependence
on $\rr$ and reduce to Eq.~\eqref{eq:r0} which in turn reduces to Eq.~\eqref{eq:massacction}.
The mass-action form of Eq.~\eqref{eq:jelem-1} reduces to Eq.~\eqref{eq:jelem},
and the flux-force relation~\eqref{eq:fluxforcemech-1} is equivalent
to Eq.~\eqref{eq:fluxforce3}. 

%\section{Fitness, selection, and thermodynamics}

\section{Fitness and selection coefficient}

\label{subsec:Fitness}

The concept of \emph{fitness} can be defined differently in different
evolutionary scenarios, and finding an appropriate definition is an
important area of research in biology and ecology~\cite{metzHowShouldWe1992,roffDefiningFitnessEvolutionary2008,orrFitnessItsRole2009,spaakDifferentMeasuresNiche2023}.
Here, we propose a definition of fitness that is suitable for autocatalytic
molecular replicators, both elementary and non-elementary.

Before proceeding, we note that one could simply define fitness as
the replication rate $\rr$ at a given point in time. However, this
definition runs into problems. For instance, for a reactor in steady
state, all non-extinct replicators have the same replication rate (the steady-state dilution rate $\dil$), while all extinct replicators
have an undefined replication rate. This makes it impossible to ask important questions, like whether higher fitness replicators
do better than lower fitness ones. At a more general level, the replication rate $\rr$
is a statistic of actual performance. It does not specify how a replicator
would perform in a new environment, as usually desired
from a fitness measure~\cite{metzHowShouldWe1992,roffDefiningFitnessEvolutionary2008,orrFitnessItsRole2009,spaakDifferentMeasuresNiche2023}.

Instead, we define the fitness $f$ of replicator $X$ via the implicit equation,
\begin{equation}
\fitness=\wwd[f]\ge0,\label{eq:fitdef}
\end{equation}
where $\wwd[\cdot]$ is the forward rate constant defined in
Eq.~\eqref{eq:effphi-1}. As we discuss below, this definition of fitness is experimentally measurable and operationally
meaningful. For an elementary replicator, $\wwd[\rr]$ does
not depend on $\rr$ and $f$ is simply the rate constant of the elementary
reaction in Eq.~\eqref{eq:massacction}. For a non-elementary replicator,
$f$ is the nonnegative root of the algebraic expression $\alpha-r(\alpha)$.
This expression is strictly increasing in $\alpha$, ranging from
$0-\wwd[0]\le0$ to $\wwd[0]-\wwd[{\wwd[0]}]\ge0$, as can be deduced
from Eqs.~\eqref{eq:r0} and \eqref{eq:derivsR}. There is a unique value of $0\le f\le\wwd[0]$
that satisfies Eq.~\eqref{eq:fitdef} and it can be found quickly using
numerical methods, such as bisection.

In operational terms, the fitness $f$ can be understood as the initial growth
rate at small concentrations. Imagine a reactor in
steady state that does not contain replicator $X$ at time $t<0$,
and suppose that $X$ is introduced at a small concentration $x(0)$
at $t=0$. Suppose also that $X$ is an elementary replicator or a
non-elementary replicator that obeys Eq.~\eqref{eq:qa} (i.e., relative
concentrations of intermediates are approximately stationary). Assuming
no dilution or degradation, the replicator's
concentration will initially grow as
\[
\dot{x}=\mathcal{J}=\rr x\approx\wwd[\rr]x,
\]
as follows from Eqs.~\eqref{eq:ggdef} and \eqref{eq:jelem-1}, while dropping second-order
terms in $x$. This implies $\rr=\wwd[\rr]$, which is uniquely satisfied
by $\rr=f$. Thus,  the concentration will initially grow as
\begin{equation}
x(t)\approx e^{t\fitness}x(0).\label{eq:iga}
\end{equation}
Moreover, an initially rare mutant with fitness $f$ will increase in concentration
if $\fitness>0$ and decrease toward extinction if $\fitness<0$. 
In
biology, this type of fitness measure is called \emph{invasion fitness}~\cite{liModelingMicrobialMetabolic2020a,arnoldiInvasionsEcologicalCommunities2022a,metzHowShouldWe1992,spaakDifferentMeasuresNiche2023}, and it has been argued to be a particularly general definition of
fitness~\cite{metzHowShouldWe1992,roffDefiningFitnessEvolutionary2008}. The  derivation above also holds for a flow reactor with dilution rate $\dil$,
in which case it gives $x(t)\approx e^{t(\fitness-\dil)}x(0)$. 

In principle, the initial growth rate is experimentally measurable. The
initial concentration $x(0)$ should be sufficiently small so that one can
neglect the reverse flux (second term in Eq.~\eqref{eq:jelem-1}) and any impact on the steady-state values of $\aa$ and $\dil$ over
the measurement timescale. At the same time, it should be sufficiently large so that stochastic
fluctuations can be ignored.

There is also another interpretation of fitness as the \emph{critical
dilution rate}, a quantity that plays an important role in chemostat
studies~\cite{pirtExtensionTheoryChemostat1970,huEngineeringPrinciplesBiotechnology2017}.
Imagine a steady-state flow reactor with concentrations
$\aa$ and steady-state dilution rate $\dilss>0$. Suppose that the
reactor contains replicator $X$ at steady-state concentration $\xxss$.
Using Eq.~\eqref{eq:jelem-1}, we can express $\xxss$ as
\begin{equation}
\xxss=\begin{cases}
\frac{\wwd[\dilss]-\dilss}{\wwdM[\dilss]} & \text{if\,}\wwd[\dilss]>\dilss\\
0 & \text{otherwise}
\end{cases},\label{eq:sol3}
\end{equation}
where we used that $\rr=\dil$ in steady state. Now imagine slowly
increasing the dilution rate $\dilss$ while maintaining constant $\aa$ (the concentrations of substrates and waste products). The replicator will be pushed to extinction at the
critical dilution rate $\dilcrit$ where $\dilcrit=\wwd[\dilcrit]$.
Given Eq.~\eqref{eq:fitdef}, the critical dilution is equal the fitness
$f$. The critical dilution rate is experimentally accessible, as
long as it is possible to increase the dilution rate while maintaining the
concentrations $\aa$ constant. 

%Interestingly, the critical
%dilution rate can be defined for hypercycles and other higher-order
%replicators for which the replication rate vanishes at small concentrations~\cite{schloglChemicalReactionModels1972a,szathmarySimpleGrowthLaws1991}.

Most simply, our definition of fitness can be considered as maximum
replication rate that can be achieved by the replicator. That is,
it is not difficult to show, for instance by using Eq.~\eqref{eq:jelem-1}, that $\rr\le f$ always. Thus,
$f$ sets an upper bound on the replication rate. This bound is approached
in the limit of low concentrations and/or high dilution rates.

In addition to fitness, we also make use of the notion of the \emph{selection
coefficient} from evolutionary biology. The selection coefficient
is a normalized measure of relative fitness difference that ranges
from 0 (no difference) to 1 (maximum difference). Given two replicators
$X$ and $X'$ with fitness values $f\ge f'$,  a common definition
of the selection coefficient is~\cite{lenski1991long}
\begin{equation}
s:=1-\frac{f'}{f}.\label{eq:selcoef}
\end{equation}

We finish by noting two important details. First, the fitness $f$ depends on the concentrations of substrates and side products $\aa$, although this is left implicit in our notation.
These concentrations may be considered as the replicator's  ecological environment.

Second, in real-world experiments, it is often difficult to measure individual
concentrations of replicator and intermediate species. Often what
is measured is a weighted sum of concentrations, 
\begin{equation}
\omega=c_{X}x+\sum_{i=1}^{m-1}c_{i}y_{i},\label{eq:comb}
\end{equation}
given some nonnegative coefficients $c_{X}$ and $c_{i}$. The interpretations
of replication rate, fitness, and critical dilution rate generalize
to this situation. That is, if the replicator and all intermediate
species grow at the same rate $\rr$, then their weighted sum $\omega$
will also grow at the same rate. Similarly, if the replicator and
all intermediate species vanish at some critical dilution rate $\hat{\dil}$,
then $\omega$ will also vanish at that dilution rate.

\section{Thermodynamic bounds}

\label{sec:darw}

To derive our thermodynamic bounds, we consider a 
replicator $X$ with a non-negative growth rate $\rr\ge0$. Our first
bound relates the affinity of replication $\ep$, the
fitness $f$, and the replication rate $\rr$ as
\begin{equation}
\ep\ge-\ln\left(1-\frac{\rr}{\fitness}\right),\label{eq:res0}
\end{equation}
which appeared as inequality~\eqref{eq:introbnd} in the Introduction.
For the special case of a flow reactor in steady state, this bound
can be written in terms of the steady-state dilution rate $\dil=\rr$
as
\begin{equation}
\epss\ge-\ln\left(1-\frac{\dil}{\fitness}\right),\label{eq:res0-1}
\end{equation}
where $\epss$ is affinity of replication under steady-state
concentrations.

To derive this bound, we combine Eqs.~\eqref{eq:ggdef} and \eqref{eq:jelem-1}
to express the replicator's concentration as $x=(\wwd[\rr]-\rr)/\wwdM[\rr]$.
Plugging into the expression of $\ep$ in Eq.~\eqref{eq:fluxforcemech-1}
gives
\begin{equation}
\ep=\ln\left(\frac{\wwd[0]}{\wwdM[0]}\frac{\wwdM[\rr]}{\wwd[\rr]-\rr}\right).\label{eq:n2}
\end{equation}
Inequality~\eqref{eq:res0} is then equivalent to $\frac{\wwd[0]}{\wwdM[0]}\frac{\wwdM[\rr]}{\wwd[\rr]-\rr}\ge f/(f-\rr)$
for $0\le\rr\le f$, which in turn is equivalent to the nonnegativity
of the function 
\begin{equation}
h(\rr):=\frac{r(0)}{r^{-}(0)}r^{-}(\rr)(f-\rr)-f(r(\rr)-\rr)\,.\label{eq:hdef}
\end{equation}
Taking second derivatives and using Eq.~\eqref{eq:derivsR} shows that $\partial_{\rr}^{2}h(\rr)\le0$,
so $h$ is concave.  Inspection shows that $h(\rr)=0$ for $\rr=0$
and $\rr=f$. Therefore, $h(\rr)\ge0$ over $0\le\rr\le f$, proving
the inequality~\eqref{eq:res0}.

Inequality~\eqref{eq:res0-1} becomes tight for elementary replicators, where the rate constants
$r(\rr)$ and $r^{-}(\rr)$ do not depend on $\rr$, so $h(\rr)=0$
for all $\rr$. More generally, our bound tends to be tighter for
``effectively'' elementary replicators with fast internal
rates, so that $r(\rr)$ and $r^{-}(\rr)$ depend weakly on $\rr$.
For general non-elementary replicators, the bound tends to be tighter
near $\rr=0$ and $\rr=f$, corresponding to equilibrium regime and absolutely irreversible
regime, respectively.

We also derive a thermodynamic bound on the selection coefficient,
which has implications for the strength of Darwinian selection. We
consider a flow reactor in steady state with dilution rate $\dil$.
Suppose that the reactor contains some replicator $X$ with fitness
$f>\dil$, and that a second replicator $X'$ with fitness
$f'\le\dil$ is pushed to extinction. Plugging into inequality
\eqref{eq:res0-1} gives
\begin{equation}
\epss\ge-\ln\left(1-\frac{f'}{f}\right)=-\ln s,\label{eq:res1}
\end{equation}
which appeared as inequality~\eqref{eq:introres} in the Introduction.
This result is illustrated on a model of replicators in a chemostat
in Section~\ref{sec:example}.

To build intuition regarding the bound~\eqref{eq:res1}, we may consider
two extreme situations. The first is equilibrium steady state, where
the replication rate $\rr$ and the affinity 
$\ep$ vanish for all replicators. All replicators are present
in positive equilibrium concentrations that do not depend on fitness,
reflecting the fact that Darwinian selection is impossible in equilibrium~\cite{eigenSelforganizationMatterEvolution1971,bernstein1983darwinian}.
At the other extreme is the irreversible regime, where each replicator
copies itself at its maximum possible rate $\rr=f$ and $\sigma$
diverges. Typically, steady states do not exist in this regime, 
since any replicator with $f>\dilss$
 grows without bound and any replicator with $f<\dil$  decays
to extinction. In the special case where a single fittest replicator satisfies
$f=\dilss$, there is a non-zero steady state containing only that replicator~\cite{kingAutocatalysis1978,lifsonCrucialStagesOrigin1997}. 
%, since a replicator with any other fitness value will go extinct. 
To
summarize, all replicators coexist in equilibrium ($\sigma=0$), while
only the fittest replicator can possibly exist in steady state in the irreversible
regime ($\sigma\to\infty$). Intermediate values of $\sigma$ interpolate
between these two extremes, permitting steady-state coexistence of
some but not all replicators.

The inequalities~\eqref{eq:res0} and~\eqref{eq:res1} are remarkably
general, being independent of most details of the chemical system.
For instance, they do not depend on the number of coexisting replicators,
the substrates/side products involved in replication, whether the
replicators copy themselves via elementary or non-elementary reactions,
whether the steady state is near or far from equilibrium, etc. They
also do not depend on the dynamical mechanism that leads to a particular
steady state. For example, they do not depend on whether replicators
experience competitive interactions (e.g., different replicators rely
on the same substrate) or not (e.g., different replicators do not
share substrates but differ in their kinetic parameters).

\section{Example: Self-complementary dimer}

\label{subsec:Example}

\begin{figure}
\ifrsb
    \newcommand{\figCwidth}{260pt}
\else
    \newcommand{\figCwidth}{1\columnwidth}
\fi
\begin{center}
\includegraphics[width=\figCwidth]{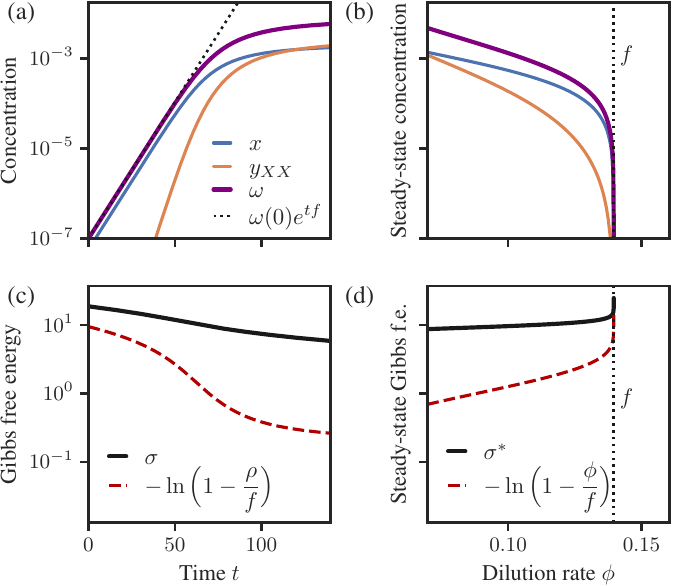}
\end{center}
\caption{\label{fig:fitness1}Fitness and thermodynamic bounds illustrated
on Rebek's self-complementary dimer~\cite{tjivikuaSelfreplicatingSystem1990a}
with 4 reactions, as in Figure~\ref{fig:complex}~(right). (a) Fitness recovers
the initial replication rate given a small starting concentration.
We show time-dependent concentrations of replicator $x$, bound dimer
$y_{XX}$, and weighted sum $\omega$ from Eq.~\eqref{eq:wsumrebek}. (b)
Fitness recovers the critical dilution rate in a flow reactor. (c) The bound~\eqref{eq:res0} relates replication rate $\protect\rr$,
fitness $f$, and affinity of replication $\protect\ep$.
It is shown on the same system as in (a). (d) The bound~\eqref{eq:res0-1}
relates fitness, steady-state affinity $\protect\epss$,
and dilution rate $\protect\dil$ in a steady-state flow reactor.
It is shown on the same system as in (b).}
\end{figure}

To illustrate our results on a concrete example, we consider
a non-elementary replicator that copies itself via the mechanism
\ifrsb
 \begin{center}
\else
 \vspace{5pt}
\fi
 \setchemfig{arrow coeff=0.6}
 \schemestart
 $X+A$
 \arrow{<=>}
 $XA$
 \arrow{<V=>[$B$][]}
 $XAB$
 \arrow{<=>}
 $XX$
 \arrow{<=>}
 $ X+X$
 \schemestop
\ifrsb
 \end{center}
\vspace{5pt}
 \else
\fi
\noindent $X$ is self-complementary
dimer while $A$ and $B$ are substrates, while the reaction $XAB\rightleftharpoons XX$
is a ligation that produces the bound dimer $XX$. This type of system
was studied in many early experiments on self-replicating chemical
systems~\cite{von1986self,zielinskiAutocatalyticSynthesisTetranucleotide1987a,tjivikuaSelfreplicatingSystem1990a,rotelloSigmoidalGrowthSelfreplicating1991,von_kiedrowski_parabolic_1991}.  It is shown schematically in Figure~\ref{fig:complex}~(right).

We parameterize the rate constants of the two binding reactions, $X+A\rightleftharpoons XA$
and $XA+B\rightleftharpoons XAB$, as 
\begin{align*}
\nu_{1}=\kappa a & \qquad\pospadA\nu_{1}^{-}=\kappa e^{\Delta G_{1}^{\circ}/RT}\\
\nu_{2}=\kappa b & \qquad\pospadA\nu_{2}^{-}=\kappa e^{\Delta G_{2}^{\circ}/RT}
\end{align*}
where $a$ and $b$ are concentrations of $A$ and $B$. The rate
constants for ligation $XAB\rightleftharpoons XX$ are parameterized as 
\begin{align*}
\nu_{3}=1\qquad & \pospadA\nu_{3}^{-}=e^{\Delta G_{3}^{\circ}/RT}
\end{align*}
and for dimerization $XX\rightleftharpoons X+X$ as 
\[
\nu_{4}=\kappa e^{-\Delta G_{4}^{\circ}/RT}\qquad\pospadA\nu_{4}^{-}=\kappa
\]

Where possible, we use parameter values from Rebek's system, one of
the first molecules that exhibited self-replication in the lab~\cite{tjivikuaSelfreplicatingSystem1990a,vonkiedrowskiMinimalReplicatorTheory1993a}.
The binding and unbinding reactions are assumed to be essentially
in equilibrium, so we use a fast rate constant $\kappa=10^{9}$. The
standard Gibbs energies for the binding reactions are $-\Delta G_{1}^{\circ}/RT=-\Delta G_{2}^{\circ}/RT=\ln60$
(favoring binding), and for dimerization it is $-\Delta G_{4}^{\circ}/RT=-\ln630$
(favoring the bound dimer)~\cite{tjivikuaSelfreplicatingSystem1990a}.
The ligation step is assumed to be highly irreversible, so we
use a large standard Gibbs energy of $-\Delta G_{3}^{\circ}/RT=10$. 

Following the experimental setup~\cite{vonkiedrowskiMinimalReplicatorTheory1993a},
we consider the weighted sum of concentrations
\begin{equation}
\omega=x+y_{XA}+y_{XAB}+2y_{XX},\label{eq:wsumrebek}
\end{equation}
which is the total concentration of replicator and intermediates,
with the bound dimer counting as two copies.

We first consider a closed reactor ($\phi=0$) which starts from nonequilibrium
initial concentrations of substrates $a(0)=b(0)=8.2\,\mathrm{mM}$~\cite{tjivikuaSelfreplicatingSystem1990a}. We choose $x(0)=.1\,\mu\mathrm{M}$
for the initial replicator concentration. For these parameter values
and substrate concentrations, we can use Eq.~\eqref{eq:effphi-1} and Eq.~\eqref{eq:fitdef}
to compute the fitness as
\[
f\approx0.14
\]
Figure~\ref{fig:fitness1}~(a) shows the time-dependent concentrations, along with predicted growth at small concentration ($\omega(0)e^{tf}$, dashed line). We see that  fitness  accurately captures the initial growth
rate. 

Next, we consider the same system, but now in steady state in a flow reactor. 
In Figure~\ref{fig:fitness1}~(b), we show steady-state concentrations
across different dilution rates, with substrate concentrations maintained
at   $a=b=8.2\,\mathrm{mM}$. We see that fitness accurately 
captures the critical dilution rate at which the replicator goes extinct.

Finally, we illustrate our thermodynamic bounds on the same system. Figure~\ref{fig:fitness1}~(c)
shows the affinity of replication versus the bound~\eqref{eq:res0}
for the system considered in Figure~\ref{fig:fitness1}~(a), where the replicator
grows from a small initial concentration in a closed reactor. The
bound is tightest in the regime of low concentrations, achieving an
efficiency of $-\ln(1-\rr/\fitness)/\ep\approx0.5$ around $x\approx.1\,\mu\mathrm{M}$.
In Figure~\ref{fig:fitness1}~(d), we show the affinity of replication
versus our bound~\eqref{eq:res0-1} for the steady-state flow reactor.
The bound is tightest near the critical dilution rate, where it achieves
an efficiency of $-\ln(1-\dil/\fitness)/\epss\approx0.65$.

It should be noted that Rebek's system, like many other self-complementary
dimers, is sometimes described as a ``parabolic replicator'' with
growth obeying a square-root law, $\dot{\omega}\propto\omega^{1/2}$.
Square-root growth results because the autocatalyst becomes bound 
in the thermodynamically-favored dimer $XX$~\cite{vonkiedrowskiMinimalReplicatorTheory1993a}.
In general, however, the type of growth varies with replicator concentration~\cite{vonkiedrowskiMinimalReplicatorTheory1993a}. At low concentrations,
the concentration of the dimerized form is small and growth is first
order, as seen in the semi-logarithmic plot Figure~\ref{fig:fitness1}~(a).
Square-root growth only appears at larger concentrations, once the
concentration of the dimerized form (orange curve in the figure) is
sufficiently large.

\section{Example: Darwinian selection in a chemostat}

\label{sec:example}

We now illustrate our thermodynamic bound on selection using a classic
model of autocatalytic replicators in a chemostat (continuous stirred-tank
flow reactor)~\cite{schusterDynamicsEvolutionaryOptimization1985}. 

The reactor may contain up to $\modelN$ replicator types, indicated
as $X_{1},\dots,X_{\modelN}$. Each $X_{i}$ undergoes an autocatalytic
reaction 
\begin{equation}
X_{i}+\MM\rightleftharpoons X_{i}+X_{i}\label{eq:auto2}
\end{equation}
from a shared substrate $A$. The substrate $\MM$ is supplied at 
concentration $\CC$ and flow rate $\chemodil$. All
species flow out with constant dilution rate $\chemodil$. 

For simplicity, we suppose that all replicators copy themselves via
elementary autocatalytic reactions. The dynamics of concentrations
of replicators $x_{i}$ and substrate $a$ are
\begin{align}
\begin{aligned}\dot{x}_{i} & =\csr x_{i}(a-e^{\GG i}x_{i})-\chemodil x_{i}\\
\dot{a} & =\chemodil(\CC-a)-\sum_{i}\csr x_{i}(a-e^{\GG i}x_{i}),
\end{aligned}
\label{eq:dyn1}
\end{align}
where $\csr$ is a rate constant and $-\GG i$ is the standard Gibbs energy of the reaction~\eqref{eq:auto2}. As usual, we leave
dependence on time of $x(t)$ and $a(t)$ implicit.

Although the replicators do not interact directly, they experience
an effective interaction due to competition for the shared substrate
$A$. This system is closely related to models of resource competition studied
in mathematical ecology and evolutionary biology~\cite{dykhuizen1983selection,smithTheoryChemostatDynamics1995,groverResourceCompetition1997,harmandChemostat2017}.
Moreover, this system can be mapped onto a competitive Lotka-Volterra
system with an effective interaction (see \appchemostat{} in the SM). 

This type of dynamical system was considered by Schuster and Sigmund~\cite{schusterDynamicsEvolutionaryOptimization1985} (see also~\cite{feistelPhysicsSelfOrganization2011}).
They showed that for any strictly positive initial conditions, there
is a unique steady state which governs the long-term behavior. This
steady state is given by a set of coupled equations,
\begin{equation}
\aaass=\CC-\sum_{i}\xxssi,\quad\pospadA\xxssi=\max\{0,e^{-\GG i}(\aaass-\chemodil/\csr)\}.\label{eq:steadyState0}
\end{equation}
In \appchemostat{} in the SM, we show how to solve the coupled equations
in Eq.~\eqref{eq:steadyState0} by evaluating at most $\modelN$ closed-form
expressions.

The strength of selection grows with increasing dilution rate $\chemodil$
and/or decreasing substrate feed concentration $\CC$, causing the
replicators to be driven to extinction one-by-one in order of increasing
$\csr$. In \appchemostat{} in the SM, we show that replicator $X_{i}$
becomes extinct once 
\begin{equation}
\frac{\CC}{\chemodil}\le\csr^{-1}+\sum_{\csrIxx:\csr[\csrIxx]\ge\csr}e^{-\GG j}(\csr^{-1}-\csr[\csrIxx]^{-1}).\label{eq:extinctionParameter}
\end{equation}

\begin{figure}[t]
\begin{center}
\ifrsb
    \newcommand{\figDwidth}{260pt}
\else
    \newcommand{\figDwidth}{1\columnwidth}
\fi
\includegraphics[width=\figDwidth]{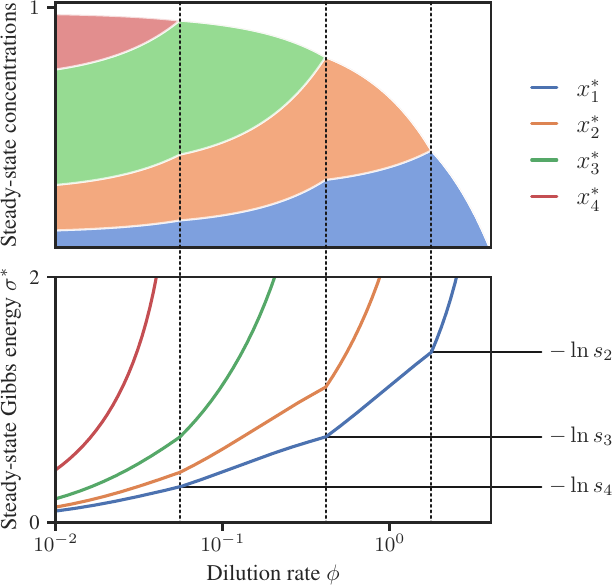}
\end{center}
\caption{\label{fig:phasetrans}Steady state behavior of a system of 4 elementary
replicators, for varying values of the dilution rate $\protect\chemodil$.
\emph{Top}: steady-state concentrations of the four replicators. As
$\protect\chemodil$ increases, the replicators are driven to extinction
one-by-one (dashed vertical lines). \emph{Bottom}: As predicted by
the bound~\eqref{eq:res1}, replicator $X_{i}$ are pushed to extinction
once the affinity of the fittest replicator (blue curve) crosses
the selection coefficient $-\ln s_{i}=-\ln(1-f_{i}/f_{1})$.}
\end{figure}

Let us consider a concrete example of 4 replicators with rate constants
$(\csr[1],\csr[2],\csr[3],\csr[4])=(4,3,2,1)$ and $-\GG{i}$
values $(1,2,3,2.5)$. Using Eq.~\eqref{eq:steadyState0}, we calculate
the steady-state concentrations $x_{i}^{*}$ of the 4 replicators at different
values of the dilution rate $\chemodil$, while substrate feed concentration
is set to $\CC=1$. The steady-state concentrations are shown in Figure~\ref{fig:phasetrans}~(top).
As the dilution rate increases, the replicators go extinct one-by-one
in order of increasing $\csr$. The critical values of $\chemodil$
at which each replicator goes extinct, as specified by Eq.~\eqref{eq:extinctionParameter},
are indicated with dotted vertical lines.

In Figure~\ref{fig:phasetrans}~(bottom), we show the steady-state affinity of each replicator, 
\begin{equation}
\epssi=\ln(\aaass/\xxssi)-\GG i.\label{eq:epss3}
\end{equation}
The values of $\epssi$ grow with increasing $\chemodil$, diverging
to infinity as each replicator approaches extinction. We compare $\sigma_{1}$,
the affinity of the fittest replicator $X_{1}$,
to the selection coefficient between $X_{1}$ and $X_{i}$, $s_{i}=1-f_{i}/f_{1}$.
Each replicator's fitness is $f_{i}=\csr\aaass$, so $s_{i}=1-\csr/\csr[1]$.
As predicted by our bound~\eqref{eq:res1}, replicator $X_{i}$ becomes
extinct once $\epssione$ crosses $-\ln s_{i}$.

We note that fitness values do not determine relative concentrations
in steady state. For instance, near equilibrium (small dilution rates),
steady-state concentrations are determined by the standard Gibbs 
energies $-\Delta G_i^{\circ}$ rather than fitness values. This can be seen in
Figure~\ref{fig:phasetrans}~(top): replicator $X_{3}$ has the largest
steady-state concentration at small $\dil$ values, since it has the
largest value of $-\Delta G_i^{\circ}$.

We can also consider the entropy production rate  
due to replication, i.e., the overall rate of dissipation of Gibbs free energy. In steady state, it is 
\begin{align}
\dot{\Sigma} & =\sum_{i}\mathcal{J}_{i}\sigma_{i}^{*}=\phi\sum_{i}x_{i}^{*}[\ln(\aaass/\xxssi)-\GG i].\label{eq:eptot}
\end{align}
Here we used Eq.~\eqref{eq:epss3} and that the steady-state flux across
the autocatalytic reaction of replicator $X_{i}$ is $\mathcal{J}_{i}=\chemodil x_{i}^{*}$.

\begin{figure}[t]
\begin{center}
\ifrsb
    \newcommand{\figEwidth}{220pt}
\else
    \newcommand{\figEwidth}{0.83\columnwidth}
\fi
\includegraphics[width=\figEwidth]{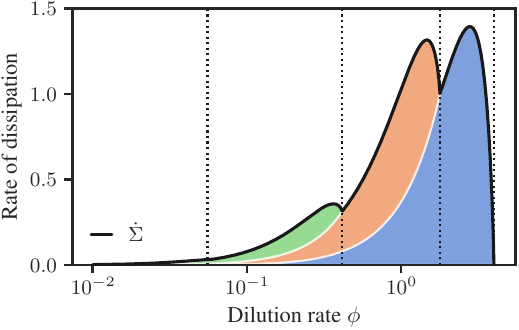}
\end{center}
\caption{\label{fig:appphasetrans-1}Black curve shows $\dot{\Sigma}$, the
overall entropy production rate, Eq.~\eqref{eq:eptot}, for
the 4-replicator model as a function of the dilution rate. Shaded
regions indicate contributions from different replicator populations,
with colors as in Figure~\ref{fig:phasetrans}. At the four extinction events
(dotted lines), the entropy production rate is continuous but not
differentiable, corresponding to second-order nonequilibrium phase
transitions.}
\end{figure}

In Figure~\ref{fig:appphasetrans-1}, we plot $\dot{\Sigma}$ for the 4-replicator
system analyzed above. Using shading, we also plot the contribution from each type
of replicator, $\phi x_{i}^{*}[\ln(\aaass/\xxssi)-\GG i]$. As before, we vary the dilution rate $\chemodil$ while holding
fixed the substrate feed concentration at $\CC=1$. Extinction events
are marked using vertical dotted lines. Recall from Figure~\ref{fig:phasetrans}
that the affinity of replication $\epss_{i}$ diverges when
replicator $X_{i}$ approaches extinction. However, the concentration
$x_{i}^{*}$ vanishes sufficiently fast so that the product $x_{i}^{*}\sigma_{i}^{*}\to0$
at extinction. As we show in \appchemostat{} in the SM, $\dot{\Sigma}$
is finite and continuous at the extinction events. Thus, under a common
classification scheme~\cite{schloglChemicalReactionModels1972a,mcneilNonequilibriumPhaseTransitions1974,zhangCriticalBehaviorEntropy2016,tomeEntropyProductionNonequilibrium2012a,nguyenExponentialVolumeDependence2020},
extinction events are second-order nonequilibrium phase transitions.

\section{Cross-catalytic cycles}

\label{sec:crosscatalytic}

We finish by briefly discussing how our results generalize to certain
types of autocatalytic sets~\cite{kauffman1986autocatalytic}. For
simplicity, we restrict our attention to autocatalytic sets with a
uniform and cyclic organization. A general treatment the thermodynamics
of cross-catalytic cycles with arbitrary topologies, kinetics, and
thermodynamic parameters is an important direction for future work.

We consider an autocatalytic set that contains $\ccNum$ species $(Z_{1},\dots,Z_{\ccNum})$
and $n$ reactions, where each species $Z_{j-1}$ catalyzes the formation of 
$Z_{j}$:
\begin{equation}
Z_{\ccIx-1}+\sum_{i}\alpha_{i}^{(\ccIx)}\MMi\rightleftharpoons Z_{\ccIx-1}+Z_{\ccIx}+\sum_{i}\beta_{i}^{(\ccIx)}\MMi.\label{eq:crosscatalyticCycle}
\end{equation}
The indexes are taken as $\modN$, so $Z_{0}=Z_{\ccNum}$, and $\alpha_{i}^{(\ccIx)},\beta_{i}^{(\ccIx)}$
indicate stoichiometric coefficients of substrates/side products participating
in each reaction. Each catalytic reaction in the cycle may be elementary,
or it may be a non-elementary mechanism as in Eq.~\eqref{eq:multistep}
but with initial reactant $X$ replaced by $Z_{\ccIx-1}$ and final
products $X+X$ replaced by $Z_{j-1}+Z_{j}$. 

\begin{figure}[b]
\ifrsb
    \newcommand{\figFwidth}{300pt}
\else
    \newcommand{\figFwidth}{1\columnwidth}
\fi
\begin{center}
\includegraphics[width=\figFwidth]{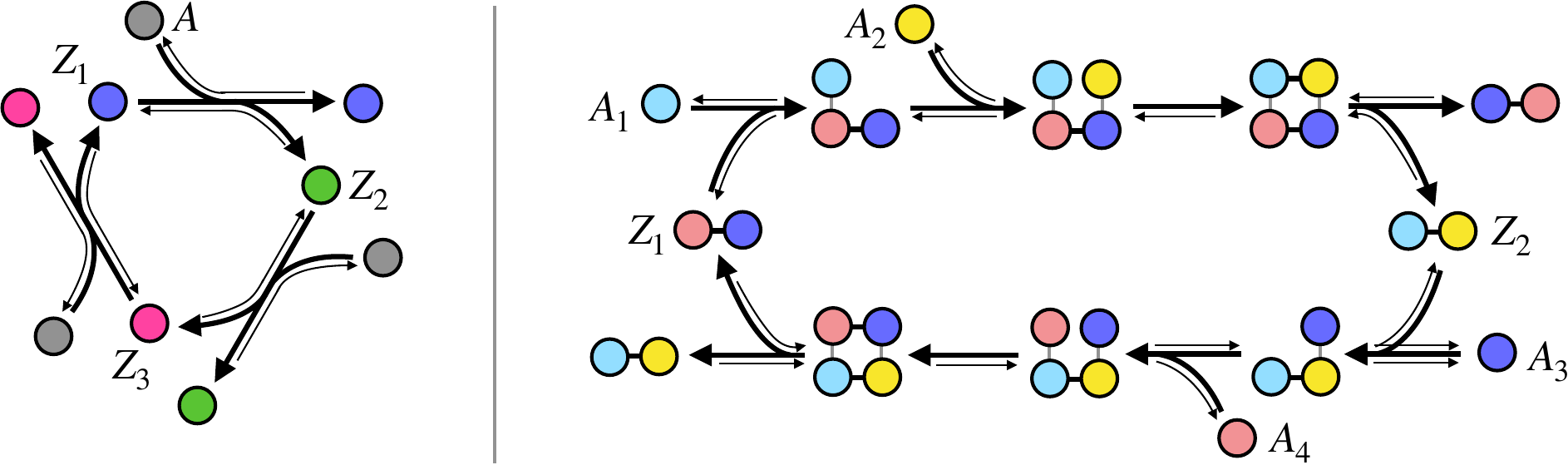}
\end{center}
\caption{\label{fig:crosscatalytic}Examples of cross-catalytic cycles. \emph{Left}:
a 3-element cycle. \emph{Right}: templated replication of complementary
polymers (shown here using dimers).}
\end{figure}

We term this kind of autocatalytic set a \emph{cross-catalytic cycle}.
A schematic illustration of a 3-member cross-catalytic cycle is shown
in Figure~\ref{fig:crosscatalytic}~(left). Cross-catalytic cycles have
attracted much attention in work on the origin-of-life, both theoretical~\cite{eigenSelforganizationMatterEvolution1971,bagleyEvolutionMetabolism1992a,hordijk2019history} and experimental~\cite{vaidyaSpontaneousNetworkFormation2012}.
An important example of a two-member cross-catalytic cycle is the
templated replication of complementary polymers, illustrated in Figure~\ref{fig:crosscatalytic}~(right),
which has been investigated in numerous experiments~\cite{sievers1994self,lincolnSelfsustainedReplicationRNA2009,bissetteMechanismsAutocatalysis2013a}.
In biology, a cross-catalytic cycle called the ``Hinshelwood cycle''
has been proposed as a general model of bacterial growth~\cite{hinshelwood136ChemicalKinetics1952,iyer2014universality}.

For simplicity, we assume that the cycle is uniform, in the sense
that each cross-catalytic reaction has the same kinetic, stoichiometric,
and thermodynamic properties. Although this assumption seems restrictive,
it suffices for studying many autocatalytic sets of fundamental interest,
such as complementary pairs with two kinetically and thermodynamically
similar cross-catalytic reactions, as in Figure~\ref{fig:crosscatalytic}~(right). 

For an autocatalytic set with a uniform cyclic organization, the cycle
members $Z_{j}$ approach equal concentrations $z_{j}\approx z_{k}$
after an initial transient. In this regime, we can effectively treat
each cycle member $Z_{j}$ as an independent replicator with autocatalytic
flux 
\begin{equation}
\mathcal{J}_{j}=\wwd[\rr]z_{j}-\wwdM[\rr]z_{j}^{2}.\label{eq:gd1}
\end{equation}
Here $\rr=\mathcal{J}_{j}/z_{j}$ is the growth rate and $\wwd[\rr],\wwdM[\rr]$
are effective forward/backward rate constants, neither of which depend
on $j$ due to the assumption of uniformity. 

The derivation of Eq.~\eqref{eq:gd1}  proceeds as follows. 
When cross-catalytic
reactions are elementary, Eq.~\eqref{eq:gd1} follows from mass-action
kinetics, $\mathcal{J}_{j}=r z_{j-1}-r^{-}z_{j}z_{j-1}$, and the assumption $z_{j}\approx z_{j-1}$. For non-elementary
cross-catalytic reactions with $m$ elementary steps, we consider
the different steps that produce or consume $Z_{j}$: the $m$-th
step of cross-catalytic reaction $j$ produces one $Z_{j}$, the first
step of cross-catalytic reaction $j+1$ consumes one $Z_{j}$, and
the $m$-th step of cross-catalytic reaction $j+1$ produces one $Z_{j}$.
The overall production of $Z_{j}$ due to the cross-catalytic cycle
is 
\begin{equation}
\mathcal{J}_{j}=J_{m}^{(j)}-J_{1}^{(j+1)}+J_{m}^{(j+1)},\label{eq:md2}
\end{equation}
where $J_{m}^{(j)}$ is the flux across the $k$-th intermediate step
in cross-catalytic reaction $j$, defined similarly to Eq.~\eqref{eq:yflux}.
Assuming uniformity of kinetics and concentrations, the intermediate
fluxes are the same for all $j$, $J_{i}^{(j)}=J_{i}^{(j+1)}\equiv J_{i}$,
therefore
\begin{align}
\mathcal{J}_{j}=2J_{m}-J_{1},
\end{align}
which recovers Eq.~\eqref{eq:netflux}. 
We can then derive Eq.~\eqref{eq:gd1} using the same analysis we used for non-elementary replicators in Section~\ref{subsec:nonelemkinetic}.

Other results follow in a similar manner as for single replicators. For instance, the affinity of each cross-catalytic reaction $j$ is $\sigmaj=\ln[\wwd[0]/z_{j}\wwdM[0]]$,
which can be shown using a similar derivation as Eq.~\eqref{eq:fluxforcemech-1}. 
The fitness $f$ of replicator $Z_{j}$ is defined via the
relation $f=\wwd[f]$. Using the same approach as in Section~\ref{subsec:Fitness}, it can be shown from Eq.~\eqref{eq:gd1} that
this fitness determines both initial growth rate and critical dilution rate.

Combining these results and using the derivation from Section~\ref{sec:darw}
gives a bound on the affinity of each cross-catalytic reaction
$j$,
\begin{equation}
\sigmaj\ge-\ln\left(1-\frac{\rr}{f}\right),\label{eq:ccbnd}
\end{equation}
thereby recovering the analogue of inequality~\eqref{eq:res0}. In steady
state with dilution rate $\dil$, this reduces to
\begin{align}
\sigmaj^{*}\ge-\ln\left(1-\frac{\dil}{f}\right),
\label{eq:ccssbnd}
\end{align}
where $\sigmaj^{*}$ is the affinity of replication
under steady-state concentrations.

We may derive a thermodynamic bound on the strength of selection
for cross-catalytic cycles, analogous to inequality~\eqref{eq:res1}. We consider
a flow reactor in steady state with dilution rate $\dil$ that contains
a cross-catalytic cycle with fitness $f>\dil$. Suppose there is another
replicator $X'$, which may be either a cross-catalytic cycle or an
individual autocatalytic reaction, that has fitness $f'\le\phi$ and
is therefore extinct in steady state. Plugging $f'\le\phi$ into
inequality~\eqref{eq:ccssbnd} gives
\begin{equation}
\sigmaj^{*}\ge-\ln\left(1-\frac{f'}{f}\right)=-\ln s.\label{eq:res1cross}
\end{equation}
This bounds the average affinity of a cross-catalytic
reaction in terms of the selection coefficient. 
%\), generalizing the bound~\eqref{eq:res1}.

Importantly, these bounds are all stated in terms of the affinity of a single reaction in the cross-catalytic cycle. For
example, for a self-replicating complementary polymer as in Figure~\ref{fig:crosscatalytic}~(right),
these inequalities bound the affinity of making 
a single complementary copy, i.e., half of the overall cycle. The
combined affinity of all $n$ reactions in the cycle obeys 
\begin{align}
\sum_j \sigmaj \ge -n\ln\left(1-\frac{\rr}{f}\right).
\label{eq:cycleEP2}
\end{align}
Since affinity is equivalent to dissipated Gibbs free energy, Eq.~\eqref{eq:cycleEP2} implies that, for a given replication rate and fitness, the thermodynamic dissipation scales linearly with cycle size.

\section{Discussion and future work}

\label{sec:Discussion}

In this paper, we uncovered a general relationship between thermodynamic affinity, 
fitness, and selection strength in molecular replicators.
This relationship was derived from the principle of local detailed
balance, which plays a central role in nonequilibrium thermodynamics.
Our results complement other recent work on fundamental stoichiometric
and thermodynamic constraints on autocatalytic replication~\cite{despons2023structural,gagrani2023polyhedral,srinivas2023characterizing,srinivas2023thermodynamics,blokhuisUniversalMotifsDiversity2020,pinero_nonequilibrium_2018}. 

Our approach has similarities to work on competitive exclusion and coexistence
theory in microbial ecology~\cite{smithTheoryChemostatDynamics1995,groverResourceCompetition1997,ellnerExpandedModernCoexistence2019}.
However, our underlying question differs from the one
typically posed in ecology: ecologists aim to explain how
diversity is maintained despite competition~\cite{ellnerExpandedModernCoexistence2019},
while we aim to explain how selection  (decrease of diversity) occurs 
despite the drive toward thermodynamic equilibrium. Nonetheless, our
work contributes to the study of autocatalytic systems from the perspective
of theoretical ecology~\cite{leeAutocatalyticNetworksTransition1997,vaidyaSpontaneousNetworkFormation2012,szilagyi2017ecology,peng2020ecological}
and evolutionary theory~\cite{vasasEvolutionGenes2012a,takeuchiEvolutionaryDynamicsRNAlike2012,mizuuchiEvolutionaryTransitionSingle2022,ametaDarwinianPropertiesTheir2021,ameta2021self}. 

Our first inequality~\eqref{eq:res0} may be compared to a thermodynamic bound on self-replicating systems derived by England~\cite{england2013statistical}. However, the two bounds make qualitatively different predictions. We refer the reader to Ref.~\cite{kolchinskyThermodynamicDissipationDoes2024} for a critical perspective of the validity of England's bound in application to autocatalytic replicators.

Our thermodynamic bound on selection~\eqref{eq:introres} is stated in terms of $\epss$, 
the affinity of replication in steady state. 
Eq.~\eqref{eq:EPdef2} implies that the steady-state affinity increases with the 
standard Gibbs energy $-\Delta G^{\circ}$ as well as the driving provided by substrates/side
products (via the term $\sum_{i}(\alpha_{i}-\beta_{i})\ln a_{i}^*$), and that it 
decreases with the replicator concentration (via the term $-\ln x^*$). 
Our results do not necessarily imply that selection strength increases with stronger 
external driving; for instance, it is not always the case that selection becomes stronger 
if replication is coupled to highly dissipative side reactions.  
Although increased driving will tend to increase the standard Gibbs energy and the contribution
from substrates/side products,
it may also decrease the affinity $\ep$ by increasing the steady-state replicator concentration $x^*$. 
The precise relationship between external driving and $\epss$ depends on the specifics 
of the chemical system, and may be an interesting topic to explore in future work. 
We also note that even if stronger driving does lead to an increase in $\epss$, this may not lead to increased selection strength because the inequality~\eqref{eq:introres} is not always tight.

We mention some other limitations and directions for future
work.  First, like many other results derived using local detailed balance, our
bound is mostly meaningful for molecular replicators that are not ``too
irreversible'', meaning that $\ep$ is not too large. For instance,
the bound~\eqref{eq:res0} can be rearranged as $\rr\le f(1-e^{-\ep})$,
which reduces to the trivial result $\rr\le f$ once $\ep$ is large (e.g., for $\sigma \ge 20$, roughly the dissipation produced by the hydrolysis of a single ATP molecule).
Our bound~\eqref{eq:introres} on the selection coefficient,  $s\ge e^{-\epss}$, also becomes weak for larger $\epss$. However, the  bound~\eqref{eq:ccbnd} for cross-catalytic cycles refers
to the affinity of a single reaction in the cycle. It 
may be applicable to highly-dissipative systems that involve large cross-catalytic cycles, as long as the individual reactions in the cycle are not too irreversible.

Another limitation is that we only consider deterministic concentrations, which is justified
when molecular counts are large and stochastic fluctuations can be
ignored. However, fluctuations cannot be ignored in small systems,
nor near extinction events when concentrations approach zero~\cite{mcneilNonequilibriumPhaseTransitions1974,nitzanFluctuationsTransitionsChemical1974}.
Future work may extend our analysis to the stochastic regime. 
%It is
%also interesting to generalize beyond the assumption of ideal solution,
%which may be violated at sufficiently high concentrations.

Third, we do not consider the effect of mutations. In general, mutations
weaken the strength of selection~\cite{eigenSelforganizationMatterEvolution1971},
therefore we expect that mutations should increase the thermodynamic
costs of selection. Future work may verify this prediction
and seek stronger bounds on the thermodynamic cost of Darwinian evolution
for replicators with mutations. The introduction of mutations leads
to other interesting questions concerning the thermodynamic cost of evolution,
such as the thermodynamic costs of finding new high-fitness replicators, 
rather than merely selecting among existing replicators. In this way, one may investigate  
the thermodynamics of ``the arrival of the fittest'',
rather than ``the survival of the fittest''~\cite{fontanaArrivalFittestTheory1994,wagnerArrivalFittestHow2015}.

Finally, our study of autocatalytic sets was restricted to the
case where reactions are organized in a single uniform cycle. Future
work may consider autocatalytic sets with more general topologies,
kinetics, and thermodynamics~\cite{jain1998autocatalytic,blokhuisUniversalMotifsDiversity2020}.
Similarly, our analysis of multi-step reaction mechanisms was restricted
to linear sequences of reactions such as Eq.~\eqref{eq:multistep}, which
may be generalized in future studies to more complex replication mechanisms.

\nocite{shivakumar1996two,charalambidesEnumerativeCombinatorics2002,schusterDynamicsEvolutionaryOptimization1985,takeuchi_existence_1980}

\newcommand{\myacks}{I thank Gülce Kardeş, Nathaniel Virgo, Jenny Poulton, David Saakian,
and Jordi Piñero for useful conversations and suggestions. This
project has received funding from the European Union’s Horizon 2020
research and innovation programme under the Marie Skłodowska-Curie
Grant Agreement No.~101068029.}
\ifrsb
    \enlargethispage{20pt}
    \ack{\myacks}
\else
    \begin{acknowledgments}
    \myacks
    \end{acknowledgments}
\fi 

\ifrsb
    \vskip2pc
    \bibliographystyle{RS} %%%% .BST file
    \bibliography{main} %%%%% .Bib file
\else
    \bibliographystyle{IEEEtran}
    \bibliography{main}
\fi

\vfill
\clearpage

%\appendix

\ifrsb
% HERE RSB 

\else

\onecolumngrid
{
    \begin{center}
        {\LARGE Thermodynamics of Darwinian selection in molecular replicators}\vspace{10pt}

        {\large Artemy Kolchinsky}\vspace{10pt}
        
        {\large \bf SUPPLEMENTARY MATERIAL}
    \end{center}
    \vspace{10pt}

}
\twocolumngrid
\renewcommand\thesection{\Alph{section}}
\counterwithout{equation}{section}
\def\theequation{S\arabic{equation}}
\def\thesubsection{\arabic{subsection}}
\makeatletter
\def\@seccntformat#1{\csname the#1\endcsname.\quad}
\makeatother
\setcounter{equation}{0}
\setcounter{section}{0}

\section{Degradation reactions}

\label{app:degradation}

Here we show that our thermodynamic bounds can be generalized to account
for degradation reactions.

Let us consider the case where the replicator $X$ undergoes degradation.
In the absence of degradation, the replicator's concentration changes
as 
\begin{equation}
\dot{x}=\mathcal{J}-\dil x=\wwd[\rr]\xx-\wwdM[\rr]\xx^{2}-\dil x,\label{eq:app0}
\end{equation}
where $\mathcal{J}=\wwd[\rr]\xx-\wwdM[\rr]\xx^{2}$ is the net flux
across the autocatalytic reaction. Now suppose that the replicator
undergoes a degradation reaction,
\begin{equation}
X+\sum_{i}\alpha_{i}'\MMi\to\sum_{i}\beta_{i}'\MMi\,,\label{eq:deg2}
\end{equation}
where $\alpha_{i}'$ and $\beta_{i}'$ are stoichiometric coefficients.
We assume that the degradation is irreversible and that the flux across
the degradation reaction is first order with rate constant $\eta\ge0$,
so $\eta x$. As usual, the rate constant $\eta$ may depend on the
concentrations of substrates/side products $\aa$, though we leave
this implicit.

In the presence of degradation, the replicator's concentration changes
as
\begin{equation}
\dot{x}=\mathcal{J}-\phi x=\wwd[\rr]x-\wwdM[\rr]x^{2}-\eta x-\phi x.\label{eq:deriv-1}
\end{equation}
We assume that $\eta<\wwd[\rr]$, since otherwise positive growth
is impossible. We then define a modified forward rate constant, 
\begin{equation}
\tilde{r}(\rr):=\wwd[\rr]-\eta\ge0,\label{eq:mod1}
\end{equation}
so that Eq.~\eqref{eq:deriv-1} becomes
\begin{equation}
\dot{x}=\tilde{r}(\rr)x-\wwdM[\rr]x^{2}-\phi x,\label{eq:deriv-1-3}
\end{equation}
thereby recovering the form of Eq.~\eqref{eq:app0}. The flux-force relation, Eq.~\eqref{eq:fluxforcemech-1} in the main text, becomes an inequality,
\begin{equation}
\ep=\ln\frac{\wwd[0]}{\xx\wwdM[0]}\ge\ln\frac{\tilde{r}(0)}{\xx\wwdM[0]},\label{eq:ff-deg}
\end{equation}
since $\tilde{r}(0)\le\wwd[0]$. We can then define fitness $\tilde{f}$
in the same way as Eq.~\eqref{eq:fitdef}, except using $\tilde{r}(\rr)$
rather than $\wwd[\rr]$. The derivation of our bounds in Section~\ref{sec:darw} in the main text 
proceeds in the same manner as before using these new definitions.
Note that the derivative relations in Eq.~\eqref{eq:derivsR} still apply
to $\tilde{r}(\rr)$, since $\eta$ does not depend on $\rr$.

Let us now consider the case of a non-elementary replicator $X$ where
some intermediate species $Y_{k}$ also undergo irreversible degradation
with rate constants $\pi_{k}\in[0,\nu_{k})$. To account for this,
the matrix $\bar{M}$ in the linear system~\eqref{eq:yvec0} turns
into
\begin{equation}
\bar{M}_{ij}:=M_{ij}+\delta_{ij}\pi_{i}.\label{eq:mmatrix-2-1}
\end{equation}
Matrix calculus shows that
\[
\frac{\partial\bar{M}_{jk}^{-1}}{\partial\bar{M}_{ii}}=-\bar{M}_{ji}^{-1}\bar{M}_{ik}^{-1}.
\]
For any set of degradation rates $\pi_{i}$, $\bar{M}$ is an M-matrix, 
so it is invertible and all entries of its inverse are nonnegative~\cite{shivakumar1996two,meyer1978singular}. Therefore, every entry
of $\bar{M}^{-1}$ decreases with increasing $\pi_{k}$. This means
that the effective rate constants $\bar{r}(\rr)$ and $\bar{r}^{-}(\rr)$,
defined as in Eq.~\eqref{eq:effphi-1} using $\bar{M}$, obey 
\[
\bar{r}(\rr)\le\wwd[\rr]\qquad\pospadA \bar{r}^{-}(\rr)\ge\wwdM[\rr].
\]
To account for the degradation of the replicator itself with rate
$\eta$, we can further modify the effective forward rate constant
as
\[
\tilde{\bar{r}}(\rr):=\bar{r}(\rr)-\eta,
\]
similarly to Eq.~\eqref{eq:mod1}. As above, we have a flux-force inequality,
\begin{align*}
\ep=\ln\frac{\wwd[0]}{\xx\wwdM[0]} & \ge\ln\frac{\bar{r}(\rr)}{\xx\bar{r}^{-}(\rr)}\ge\ln\frac{\tilde{\bar{r}}(\rr)}{\xx\bar{r}^{-}(\rr)}.
\end{align*}
The derivation of our bounds then proceeds as before, after defining
fitness $\tilde{\bar{f}}$ using the forward rate constant $\tilde{\bar{r}}(\rr)$.
Note that the derivative relations in Eq.~\eqref{eq:derivsR} still apply
to $\tilde{\bar{r}}(\rr)$, since $\bar{M}$ is an M-matrix and $\eta$
does not depend on $\rr$.

In general, our thermodynamic bounds will tend to be less tight in
the presence of degradation, since the flux-force equality~\eqref{eq:fluxforcemech-1}
turns into an inequality. At the same time, the affinity 
of replication may increase due to degradation, for instance because
the steady-state concentration of replicator decreases due to degradation.
Some of our bounds may also be larger, for instance the right side
of the steady-state bound~\eqref{eq:res0-1} will be larger, since
fitness will decrease due to degradation.

\section{Non-elementary replicators}

\label{app:multistep2}

Here we derive closed-form expressions of the effective rate constants
in the $\rr=0$ regime, as appear in Eq.~\eqref{eq:r0}. As discussed near 
the end of Section~\ref{subsec:nonelemkinetic} in the main text, this regime applies when
the replication rate $\rr$ is much slower than the rate of internal reactions.

Plugging $\rr=0$ into Eq.~\eqref{eq:QSS2} implies that $\JJ[\msIx]=\JJ[\msIx+1]$,
so all intermediate fluxes are equal:
\begin{equation}
\JJ[\msIx]=J\qquad\text{for }\msIx\in{1..\msNum}.\label{eq:alleq}
\end{equation}
They also all equal to the rate of production of the replicator, $\mathcal{J}=2\JJ[\msNum]-\JJ[1]=J$.
We can then rearrange Eq.~\eqref{eq:yflux} to write the steady-state concentrations
of $Y_{\msIx}$ as
\[
y_{\msIx}=(\rrArg{\msIx}/\rrMArg{\msIx})y_{\msIx-1}-\mathcal{J}/\rrMArg{\msIx}\qquad\msIx\in{1..\msNum},
\]
where we use the convention $y_{0}=x$, $y_{\msNum}=x^{2}$. This
is a first-order linear recurrence relation for $y_{\msIx}$. Using
an existing result~\cite[Thm. 7.1,][]{charalambidesEnumerativeCombinatorics2002},
this recurrence can be solved for $y_{\msNum}=x^{2}$ starting from
the initial condition $y_{0}=x$ to give
\begin{align}
x^{2} & =\left(x-\mathcal{J}\sum_{\msIx=1}^{\msNum}\frac{\prod_{\msIxx=1}^{\msIx-1}\rrMArg{\msIxx}}{\prod_{\msIxx=1}^{\msIx}\rrArg{\msIxx}}\right)\prod_{\msIx=1}^{\msNum}\frac{\rrArg{\msIx}}{\rrMArg{\msIx}}\label{eq:hg3}
\end{align}
where we use the convention $\prod_{\msIxx=1}^{\msIx-1}\rrMArg{\msIxx}=1$
for $k=1$. Plugging in $\wwd[0]$ and $\wwdM[0]$ from Eq.~\eqref{eq:r0}
gives 
\[
x^{2}=\left(x-\mathcal{J}/\wwd[0]\right)\frac{\wwd[0]}{\wwdM[0]}.
\]
This can be rearranged as $\mathcal{J}=\wwd[0]x-\wwdM[0]x^{2}$, which
recovers the mass-action-like form of Eq.~\eqref{eq:jelem-1}.

\section{Chemostat model}

\label{app:chemostat}

\subsection{Steady state}

\label{app:chemostat-ss}
\global\long\def\Rmatrix{R}%

Here we analyze the steady-state behavior of the dynamical system
described by Eq.~\eqref{eq:dyn1}. To begin, let $\omega:=\mm+\sum_{i}x_{i}$
indicate the total concentration of substrate and replicators at time
$t$. The first line of Eq.~\eqref{eq:dyn1} means that $\csr x_{i}[a-e^{\GG i}x_{i}]=\dot{x}_{i}+\chemodil x_{i}$.
Plugging this into the second line of Eq.~\eqref{eq:dyn1} and rearranging
gives
\[
\dot{\omega}=\chemodil(\CC-\omega).
\]
Thus, $\omega$ converges exponentially fast to the steady-state value
$\omega^{*}=\CC$. 

We consider the long-term dynamics of the system restricted to the
invariant subspace $\omega=\CC$. Within this subspace, we can rewrite
the first line of Eq.~\eqref{eq:dyn1} as 
\begin{equation}
\dot{x}_{i}=\csr x_{i}[\CC-\sum_{\csrIxx}x_{\csrIxx}-e^{\GG i}x_{i}]-\chemodil x_{i}.\label{eq:diff1}
\end{equation}
Using an appropriate Lyapunov function, Schuster and Sigmund demonstrated
that the dynamics in Eq.~\eqref{eq:diff1} converge to the steady state
in Eq.~\eqref{eq:steadyState0} for any strictly positive initial condition
$(x_{1}(0),\dots,x_{\modelN}(0))\in\mathbb{R}_{+}^{\modelN}$~\cite{schusterDynamicsEvolutionaryOptimization1985}.
A similar global convergence result can also be derived from the theory
of Lotka-Volterra dynamics~\cite{takeuchi_existence_1980}. Specifically,
Eq.~\eqref{eq:diff1} can be written as a competitive Lotka-Volterra system,
\begin{equation}
\dot{x}=b_{i}x_{i}+\sum_{\csrIxx}\Rmatrix_{i\csrIxx}x_{i}x_{\csrIxx},\label{eq:LK0-1}
\end{equation}
where $b_{i}:=\csr\CC-\chemodil$ and $\Rmatrix_{i\csrIxx}=-\csr(1+\delta_{i\csrIxx}e^{\GG i})$.
The matrix $\Rmatrix$ can be expressed as $\Rmatrix=-K(\vec{1}\vec{1}^{T}+D)$,
where $K_{i\csrIxx}=\delta_{i\csrIxx}\csr$ and $D=\delta_{i\csrIxx}e^{\GG i}$
are diagonal matrices and $\vec{1}$ is a vector of all 1s. Note that
$\vec{1}\vec{1}^{T}+D$ is positive definite, since $\vec{1}\vec{1}^{T}$
is positive semidefinite and $D$ is positive definite. It is known
that for this type of Lotka-Volterra system, any strictly positive
initial condition converges to a unique globally attracting fixed
point~\cite{takeuchi_existence_1980}, which is the steady state
specified by Eq.~\eqref{eq:steadyState0}.

The steady state in Eq.~\eqref{eq:steadyState0} is expressed as a set of
coupled equations, which can be solved in the following manner. Assume
without loss of generality that the rate constants $\csr$ are arranged
in decreasing order, $\csr[1]\ge\csr[2]\ge\dots\ge\csr[\modelN]$.
Given Eq.~\eqref{eq:steadyState0}, it must then be that $x_{i}^{*}=0$
implies $x_{\csrIxx}^{*}=0$ whenever $\csrIxx>i$. Suppose for the
moment that the top $i\in\{0..\modelN\}$ replicators have non-zero
steady-state concentrations, 
\begin{equation}
\xxss_{\csrIxx}=\begin{cases}
e^{-\GG j}(\aaass-\chemodil/\csr[\csrIxx]) & \csrIxx\le i\\
0 & \csrIxx>i
\end{cases}\label{eq:a2b}
\end{equation}
Eq.~\eqref{eq:steadyState0} then gives $\aaass=\CC-\sum_{\csrIxx=1}^{i}e^{-\GG j}(\aaass-\chemodil/\csr[\csrIxx])$,
so
\begin{align}
\aaass & =\frac{\CC+\chemodil\sum_{\csrIxx=1}^{i}e^{-\GG j}\csr[\csrIxx]^{-1}}{1+\sum_{\csrIxx=1}^{i}e^{-\GG j}}.\label{eq:a2a}
\end{align}
Eqs.~\eqref{eq:a2b} and \eqref{eq:a2a} are the solution to Eq.~\eqref{eq:steadyState0} if
for all $\csrIxx\in\{0..\modelN\}$,
\begin{equation}
x_{\csrIxx}^{*}=\max\{0,e^{-\GG j}(\aaass-\chemodil/\csr[\csrIxx])\}.\label{eq:vcn2}
\end{equation}
Given Eq.~\eqref{eq:a2b}, Eq.~\eqref{eq:vcn2} is satisfied once 
\begin{equation}
\aaass-\chemodil/\csr\ge0\ge\aaass-\chemodil/\csr[i+1],\label{eq:matc2s}
\end{equation}
Therefore, to solve Eq.~\eqref{eq:steadyState0}, it suffices to evaluate
Eqs.~\eqref{eq:a2a} and \eqref{eq:a2b} for $i=0,1,2,\dots$, stopping once the bounds
\eqref{eq:matc2s} are satisfied.

\subsection{Derivation of inequality~\eqref{eq:extinctionParameter} (condition
for extinction)}

Given Eq.~\eqref{eq:steadyState0}, replicator $X_{i}$ is extinct once
\begin{equation}
\aaass\le\chemodil/\csr.\label{eq:m3}
\end{equation}
If this inequality holds, then any lower fitness replicator $X_{\csrIxx}$
($\csr[\csrIxx]\le\csr$) must also be extinct, since then $a\le\chemodil/\csr[\csrIxx]$.
Now, combine the equations in Eq.~\eqref{eq:steadyState0} to write
\begin{align}
\aaass & =\gamma-\sum_{\csrIxx:x_{\csrIxx}>0}e^{-\GG j}(\aaass-\chemodil/\csr[\csrIxx])\nonumber \\
 & \le\gamma-\sum_{\csrIxx:\csr[\csrIxx]>\csr}e^{-\GG j}(\aaass-\chemodil/\csr[\csrIxx]),\label{eq:df2}
\end{align}
where the inequality in the second line reflects that it may be that
$\aaass\le\chemodil/\csr[\csrIxx]$ even for higher fitness replicators
($\csr[\csrIxx]>\csr$). Rearranging inequality~\eqref{eq:df2} gives
\begin{equation}
a\le\frac{\gamma+\chemodil\sum_{\csrIxx:\csr[\csrIxx]\ge\csr}e^{-\GG j}\csr[\csrIxx]^{-1}}{1+\sum_{\csrIxx:\csr[\csrIxx]\ge\csr}e^{-\GG j}}.\label{eq:m4}
\end{equation}
Given inequality~\eqref{eq:m4}, the bound~\eqref{eq:m3} must be
satisfied when
\[
\frac{\gamma+\chemodil\sum_{\csrIxx:\csr[\csrIxx]\ge\csr}e^{-\GG j}\csr[\csrIxx]^{-1}}{1+\sum_{\csrIxx:\csr[\csrIxx]\ge\csr}e^{-\GG j}}\le\chemodil/\csr.
\]
Rearranging gives the inequality~\eqref{eq:extinctionParameter}.

\subsection{Nonequilibrium phase transitions at extinctions}

Here we show that the entropy production rate,
Eq.~\eqref{eq:eptot}, is continuous but not differentiable at extinction
events, meaning that extinctions are second-order nonequilibrium phase
transitions.

\global\long\def\dilcritI{\hat{\phi}_{(i)}}%
Suppose that the rate constants are strictly ordered as 
\begin{equation}
\csr[1]>\csr[2]>\dots>\csr[\modelN].\label{eq:sOrd}
\end{equation}
From Eq.~\eqref{eq:extinctionParameter}, the critical dilution rate for
replicator $X_{i}$ is
\begin{align}
\dilcritI & =\frac{\CC}{\csr^{-1}+\sum_{\csrIxx=1}^{i-1}e^{-\GG j}(\csr^{-1}-\csr[\csrIxx]^{-1})}\label{eq:extc4}\\
 & =\frac{\CC}{\csr^{-1}+\sum_{\csrIxx=1}^{i}e^{-\GG j}(\csr^{-1}-\csr[\csrIxx]^{-1})}\label{eq:extc5}
\end{align}
where in the second line we used $\csr^{-1}-\csr[\csrIxx]^{-1}=0$
for $j=i$. We treat $\chemodil$ as the control parameter, while
holding $\CC$ fixed.

We show that the steady-state substrate concentration $\aaass$ is
continuous as a function of the dilution rate at $\dilcritI$. When
$\chemodil<\dilcritI$, replicators $X_{1},\dots,X_{i}$ are not extinct.
We consider the limit from below,
\begin{align*}
\lim_{\chemodil\nearrow\dilcritI}\aaass & =\frac{\CC+\dilcritI\sum_{\csrIxx=1}^{i}e^{-\GG j}\csr[\csrIxx]^{-1}}{1+\sum_{\csrIxx=1}^{i}e^{-\GG j}}=\dilcritI\csr[i]^{-1}.
\end{align*}
Here we first used Eq.~\eqref{eq:a2a} and then plugged in Eq.~\eqref{eq:extc4}
and simplified using a bit of tedious algebra. When $\chemodil>\dilcritI$,
replicators $X_{1},\dots,X_{i-1}$ are not extinct, but replicator
$X_{i}$ is extinct. We consider the limit from above,
\begin{align*}
\lim_{\chemodil\searrow\dilcritI}\aaass & =\frac{\CC+\chemodil\sum_{\csrIxx=1}^{i-1}e^{-\GG j}\csr[\csrIxx]^{-1}}{1+\sum_{\csrIxx=1}^{i-1}e^{-\GG j}}=\dilcritI\csr[i]^{-1},
\end{align*}
where we first used Eq.~\eqref{eq:a2a} and then plugged in Eq.~\eqref{eq:extc5}
and simplified. The two limits match, so $\aaass$ is continuous at
$\dilcritI$. This also implies that steady-state replicator concentrations
$x_{\csrIxx}^{*}=\max\{0,e^{-\GG j}(\aaass-\chemodil/\csr[\csrIxx])\}$
from Eq.~\eqref{eq:steadyState0} are continuous at $\dilcritI$. 

Next, consider the entropy production rate at the critical
point,
\ifrsb
 \begin{align*}
\lim_{\chemodil\to\dilcritI}\dot{\Sigma}=\lim_{\chemodil\to\dilcritI}\chemodil\Bigg[\sum_{\csrIxx=1}^{i-1}x_{\csrIxx}^{*}\Big(\ln\frac{\aaass}{x_{\csrIxx}^{*}}-\GGfrac j\Big)+x_{i}^{*}(\ln\aaass-\GGfrac i)\Bigg],
 \end{align*}
\else
 \begin{multline*}
\lim_{\chemodil\to\dilcritI}\dot{\Sigma}=\\
\lim_{\chemodil\to\dilcritI}\chemodil\Bigg[\sum_{\csrIxx=1}^{i-1}x_{\csrIxx}^{*}\Big(\ln\frac{\aaass}{x_{\csrIxx}^{*}}-\GGfrac j\Big)+x_{i}^{*}\Big(\ln\aaass-\GGfrac i\Big)\Bigg],
 \end{multline*}
\fi
where we used $\lim_{\chemodil\to\dilcritI}x_{i}^{*}\ln x_{i}^{*}=\lim_{\alpha\to0}\alpha\ln\alpha=0.$
Given Eq.~\eqref{eq:sOrd}, $x_{\csrIxx}^{*}>0$ for $\csrIxx\in\{1..i-1\}$,
so all the terms on the right side are finite and continuous at $\chemodil=\dilcritI$.
Thus, $\dot{\Sigma}$ is a continuous function of $\chemodil$ at
$\dilcritI$.

Next, we show that $\dot{\Sigma}$ is not differentiable with respect
to $\chemodil$ at $\dilcritI$ because it has an infinite left derivative
at this point (it is not hard to show that the right derivative is
finite). The left derivative is
\begin{align*}
\partial_{\chemodil}^{-}\dot{\Sigma} & =\sum_{\csrIxx=1}^{i}\Bigg[\partial_{\chemodil}^{-}x_{\csrIxx}^{*}\Big(\ln\aaass-\ln x_{\csrIxx}^{*}-1-\GGfrac j\Big)+\frac{x_{\csrIxx}^{*}}{a^{*}}\partial_{\chemodil}^{-}a^{*}\Bigg].
\end{align*}
All terms in this expression are finite except for $-(\partial_{\chemodil}^{-}x_{i}^{*})\ln x_{i}^{*}$.
Plugging in Eq.~\eqref{eq:steadyState0} gives
\begin{align*}
\partial_{\chemodil}^{-}x_{i}^{*} & =e^{-\GG i}(\partial_{\dilss}^{-}a^{*}-\csr[i]^{-1})\\
 & =e^{-\GG i}\left[\frac{\sum_{\csrIxx=1}^{i}e^{-\GG j}\csr[\csrIxx]^{-1}}{1+\sum_{\csrIxx=1}^{i}e^{-\GG j}}-\csr[i]^{-1}\right]\\
 & =e^{-\GG i}\left[\frac{\sum_{\csrIxx=1}^{i}e^{-\GG j}(\csr[\csrIxx]^{-1}-\csr[i]^{-1})-\csr[i]^{-1}}{1+\sum_{\csrIxx=1}^{i}e^{-\GG j}}\right]\\
 &<0,
\end{align*}
where in the second line we used Eq.~\eqref{eq:a2a}, and in the last line
we used that $\csr[\csrIxx]>\csr[i]$ and $\csr[i]>0$. Hence, $\partial_{\chemodil}^{-}x_{i}^{*}$
is a strictly negative constant, meaning that $\chemodil\nearrow\dilcritI$,
and $x_{i}^{*}\to0$, $-(\partial_{\chemodil}^{-}x_{i}^{*})\ln x_{i}^{*}\to-\infty$.
This shows that the left derivative of $\dot{\Sigma}$ is negative
infinite at $\chemodil=\dilcritI$. 
\fi

\vfill

\end{document}